\documentclass[9pt,twocolumn,twoside]{pnas-new_JHEP}

\makeatletter
\newcommand{\manuallabel}[2]{\def\@currentlabel{#2}\label{#1}}
\makeatother

\setboolean{displaywatermark}{false}
\templatetype{pnasresearcharticle} 
\usepackage{amsfonts}
\usepackage{upgreek}
\usepackage{slashed}
\usepackage{amsmath,amssymb,bm,bbm} 
\usepackage{latexsym}
\usepackage[export]{adjustbox}
\usepackage[colorlinks=true]{hyperref}  
\usepackage[toc,page]{appendix}
\usepackage{dsfont}
\usepackage{pdfpages}
\usepackage{graphicx}
\usepackage{subfig}
\newcommand{\beq}{\begin{equation}}
\newcommand{\eeq}{\end{equation}}
\def\bea{\begin{eqnarray}}
\def\eea{\end{eqnarray}}
\newcommand{\nn}{\nonumber \\}

\newcommand{\x}{\bm{x}}

\usepackage{xr}
\makeatletter
\newcommand*{\addFileDependency}[1]{
 \typeout{(#1)}
 \@addtofilelist{#1}
  \IfFileExists{#1}{}{\typeout{No file #1.}}
}
\makeatother

\newcommand*{\myexternaldocument}[1]{
    \externaldocument{#1}
    \addFileDependency{#1.tex}
    \addFileDependency{#1.aux}
}
\myexternaldocument{SI2}
\title{Localization of overdamped bosonic modes and transport in strange metals}

\author[a]{Aavishkar A. Patel\textsuperscript{1}}
\author[b]{Peter Lunts}
\author[b]{Subir Sachdev}

\affil[a]{Center for Computational Quantum Physics, Flatiron Institute, New York, New York, 10010, USA}
\affil[b]{Department of Physics, Harvard University, Cambridge MA-02138, USA}

\leadauthor{Patel} 

\significancestatement{
Almost all higher temperature superconductor materials exhibit a `strange metal' regime above the critical temperature for superconductivity. The important problem of theoretically computing the critical temperature for superconductivity therefore requires a complete theory of the strange metal. We investigate the subtle consequences of multi-electron quantum entanglement in the presence of impurities at spatially random positions in a strange metal. Using modern computer hardware, we are able to identify two distinct regimes: one previously studied regime in which the impurities can be treated in an averaged manner, and the other regime in which rare impurity configurations dominate. Our results lead to a deeper understanding of the global phase diagrams of higher temperature superconductors.
}

\correspondingauthor{\textsuperscript{1}To whom correspondence should be addressed. E-mail: apatel@flatironinstitute.org}


\begin{abstract}
A recent theory described strange metal behavior in a model of a Fermi surface coupled a two-dimensional quantum critical bosonic field with a spatially random Yukawa coupling. With the assumption of self-averaging randomness, similar to that in the Sachdev-Ye-Kitaev model, numerous observed properties of a strange metal were obtained for wide range of intermediate temperatures, including the linear-in-temperature resistivity. The Harris criterion implies that spatial fluctuations in the local position of the critical point must dominate at lower temperatures.
For an $M$-component boson with $M \geq 2$,
we use multiple graphics processing units (GPUs) to compute the real frequency spectrum of the boson propagator in a self-consistent mean-field treatment of the boson self-interactions, but an exact treatment of multiple realizations of the spatial randomness from the random boson mass. We find that Landau damping from the fermions leads to the emergence of the physics of the random transverse-field Ising model at low temperatures, as has been proposed by Hoyos, Kotabage, and Vojta. This regime is controlled by localized overdamped eigenmodes of the bosonic scalar field, also has a resistivity which is nearly linear-in-temperature, and extends into a `quantum critical phase' away from the quantum critical point, as observed in several cuprates. For the $M = 1$ Ising scalar, the mean-field treatment is not applicable, and so we use Hybrid Monte Carlo simulations running on multiple GPUs; we find a rounded transition and localization physics, with strange metal behavior in an extended region around the transition.  
\end{abstract}

\dates{This manuscript was compiled on \today}
\doi{\url{www.pnas.org/cgi/doi/10.1073/pnas.XXXXXXXXXX}}

\begin{document}

\maketitle
\ifthenelse{\boolean{shortarticle}}{\ifthenelse{\boolean{singlecolumn}}{\abscontentformatted}{\abscontent}}{}
\noindent
\href{https://arxiv.org/abs/2312.06751}{\large\bf arXiv:2312.06751}



\dropcap{S}trange metals are an unusual state of quantum matter invariably present above the critical temperature of correlated electron superconductors, including the cuprate high temperature superconductors \cite{Hartnoll:2021ydi}. They are characterized by numerous properties which deviate from the Fermi liquid description of conventional metals: most prominent among these are the linear-in-temperature resistivity, and the $1/\omega$ tail in the optical conductivity \cite{Michon22}, where $\omega$ is frequency.

A recent work \cite{Patel:2022gdh}  proposed a universal theory of strange metals by considering the influence of spatially random electron-electron interactions on the theory of quantum phase transitions in metals \cite{VojtaRMP}. The spatial randomness was treated in a self-averaging manner, similar to the methods employed in the solution of the infinite-range Sachdev-Ye-Kitaev (SYK) models \cite{Chowdhury:2021qpy}. This universal theory was found to be a good description of observations in a widening fan of temperatures emerging from the zero temperature quantum critical point (QCP). 

However, strange metal behavior is often observed over wider regions of the phase diagram, and can appear in an extended region at low temperatures ($T$) away from the $T=0$ QCP \cite{Hussey_foot,Greene_rev}. Bashan {\it et al.\/} \cite{Bashan23} postulated a non-zero density of two-level systems which resonantly scatter electrons, and argued that they can led to the needed extended quantum critical phase at low $T$. 
Here, we show that the self-averaging assumed in the universal theory \cite{Patel:2022gdh} breaks down at very low $T$, and there is eventually a crossover to a regime where the overdamped bosonic modes of the quantum critical theory spatially localize. These localized bosonic modes are the analog of the two-level systems of Bashan {\it et al.\/} \cite{Bashan23}, and lead to an extended low $T$ quantum critical phase with a nearly linear in $T$ resistivity. We emphasize that our localized, overdamped bosonic modes are not postulated degrees of freedom, but emerge naturally in the existing theory after the flow to strong disorder at low $T$ is accounted for. We also note earlier works \cite{Milica89,BhattFisher,Vlad05a,Vlad05b,Vlad07} in which localized bosonic modes played an important role in other metallic correlated electron systems. 

Section~\ref{sec:bos} describes an effective theory for the bosonic modes alone, and we present numerical results on its properties for $M \geq 2$ boson flavors in Section~\ref{sec:numbos}. The derivation of the effective boson only theory from the original fermion-boson theory of Ref.~\cite{Patel:2022gdh} is presented in Section~\ref{sec:fermion}, along with a description of electrical transport. We treat the $M=1$ Ising scalar case separately using an alternative technique in Section~\ref{sec:scalar}.

\section{Effective theory for overdamped bosonic modes}
\label{sec:bos}

We begin by considering a spatially random version of the Hertz-Millis theory of metallic QCPs \cite{Vojta_rev} for a $M$-component bosonic field $\phi_{ja} (\tau)$ with action $\mathcal{S}_\phi + \mathcal{S}_{\phi d}$:
\begin{align}
    \mathcal{S}_\phi =& \int d \tau \Bigl[ \frac{J}{2}\sum_{\langle i j \rangle} \sum_a \left( \phi_{i a} -\phi_{ja} \right)^2 \nonumber \\
    & ~~~~~~~~~~~~~~+  \sum_{j} \bigl( \frac{\lambda+ \lambda'_j}{2} \sum_a \phi_{j a}^2   + \frac{u}{4M} \left( \sum_a \phi_{j a}^2 \right)^2 \bigr)\Bigr] \nonumber \\
    \mathcal{S}_{\phi d} =& \frac{T}{2} \sum_{\Omega} \sum_j \sum_a \left(\gamma |\Omega| + \Omega^2/c^2 \right)|\phi_{ja} (i\Omega)|^2 \,, \label{Sphi}
\end{align}
in two spatial dimensions. Here $a = 1 \ldots M$, $j$ labels the sites of a square lattice,  $\tau$ is imaginary time, and $\Omega$ is Matsubara frequency. The action $\mathcal{S}_\phi+ \mathcal{S}_{\phi d}$ can obtained by integrating out the fermions from the universal theory of Patel {\it et al.\/} \cite{Patel:2022gdh} (see Eq.~[\ref{S}]). All couplings in $\mathcal{S}_\phi + \mathcal{S}_{\phi d}$ would then be spatially random, but we have only retained the spatial randomness in the tuning parameter $\lambda$ because that is the most relevant form of disorder by the Harris criterion \cite{Patel:2022gdh,Vojta_rev}---we have included a spatially random $\lambda_j^\prime$, whose disorder averages obey
\begin{align}
    \overline {\lambda_j^\prime} = 0 \quad, \quad  \overline {\lambda_j^\prime \lambda_k^\prime} = \delta_{jk} \lambda^{\prime 2}\,.
\end{align}
The coupling $\gamma>0$ in $\mathcal{S}_{\phi d}$ is the Landau damping induced by the fermions, and the repulsive self-interaction $u>0$ plays an important role in stabilizing the theory in regions where $\lambda_j$ is very negative. 

The bosonic field $\phi$ can represent a symmetry-breaking order parameter (such as a spin density wave or Ising nematic order), or a fractionalized field (such as a hybridization boson in Fermi volume changing transitions \cite{DCSS15b,Aldape22,Sachdev:2023fim}). 

For the case of $M \geq 2$ the global O($M$) symmetry is continuous, and important results for the properties of $\mathcal{S}_\phi+ \mathcal{S}_{\phi d}$ were obtained by Hoyos {\it et al.\/} \cite{Hoyos07,Hoyos09}, building on earlier work \cite{CastroNeto1,Schmalian02,Schmalian05}. Assuming that the renormalization group flow of $\lambda_j^\prime$ (and also of a spatially random addition to $J$) was towards a broad distribution, Hoyos {\it et al.\/} reached the remarkable conclusion that the low temperature properties near the QCP $\lambda=\lambda_c$ were the same as that of the random transverse field Ising model for insulators {\it without\/} any fermion-induced Landau damping. This random Ising model was shown to be described by infinite randomness fixed points by Fisher \cite{DSF92,DSF95} in $d=1$ spatial dimension; numerical studies in $d=2$ \cite{Motrunich00, Pandey23} also support infinite randomness fixed points. The argument of Hoyos {\it et al.} relies on the fact that the $|\Omega|$ Landau damping term in $\mathcal{S}_{\phi d}$ is equivalent to a long-range $\phi$-$\phi$ interaction which decays as $1/\tau^2$ in imaginary time. 
Then the quantum dynamics of a droplet which is nearly ordered can be mapped on to the statistical mechanics of a one-dimensional chain of O($M$) spins with an inverse-square interaction; this classical model has no phase transition, but a correlation length which diverges exponentially with the inverse classical `temperature' \cite{Kosterlitz76}. This exponentially large correlation length is similar to the exponentially large correlation length of classical Ising chains with short-range interactions \cite{Ising25}, and hence the mapping of Hoyos {\it et al.\/}. This mapping can be understood as a compensation between the enhancement of local quantum fluctuations in a droplet by the continuous symmetry, and the suppression of quantum fluctuations by the fermion-induced dissipation, leading to a mapping to a non-dissipative quantum system with a discrete symmetry. For the case of $M = 1$, the ordered droplets have no additional fluctuations, leading to the destruction of the Griffiths phase and a smearing of the QCP \cite{VojtaIsing1,VojtaIsing2}.  

\section{Numerical results for the bosonic theory}
\label{sec:numbos}

We have obtained numerical results for the properties of $\mathcal{S}_\phi + \mathcal{S}_{\phi d}$. When $M \geq 2$, we follow the analysis of Del Maestro {\it et al.\/} \cite{Adrian08} for the same theory in a different context in $d=1$. We solve the large $M$ saddle-point equations in a finite sample exactly in the presence of specific realizations of disorder. Modern computing hardware makes such a numerical solution possible for large systems in $d=2$, as we will describe below. The large $M$ method makes it possible to perform exact analytic continuation to real time, and that will be important for our computation of the fermion spectrum and transport properties. 
Moreover, we expect the large $M$ method to accurately capture the physics of the infinite disorder physics because the exponential divergence of the droplet correlation time is also present in the classical limit. We note that the numerical large $M$ solutions in $d=1$ were in excellent agreement \cite{Adrian08} with exact theoretical predictions for the critical properties \cite{DSF92,DSF95}.

The large $M$ saddle point equations amount to replacing $\mathcal{S}_\phi$ by
\begin{align}
    \overline{\mathcal{S}}_\phi =& \int d \tau \Bigl[ \frac{J}{2}\sum_{\langle i j \rangle} \left( \phi_{i a} -\phi_{ja} \right)^2 +  \sum_{j} \frac{ \overline{\lambda}'_j}{2} \phi_{j a}^2   \Bigr] \label{Sphimean} \\
    & \overline{\lambda}'_j = \lambda + \lambda'_j + \frac{u}{M}
    \sum_a \left\langle \phi_{ja}^2 \right\rangle_{\overline{\mathcal{S}}_\phi + \mathcal{S}_{\phi d}}
    \label{renmass}
\end{align}
The numerical solution of this equation involves diagonalization of the boson propagator implied by $\overline{\mathcal{S}}_\phi$. 
At a given $\lambda$, we then compute self-consistent renormalized `mass' $\overline{\lambda}_j$ by updating it iteratively using Eq.~[\ref{renmass}].

We used $J = 1,~\gamma = 1,~c^2 = 10$, and $u=1$ throughout this work, with the variance of the random mass, {\it i.e.} ${\lambda'}^2$, taken to be $0.25$. The size of the $L\times L$ systems was set to $L = 160$ unless otherwise mentioned, with periodic boundary conditions. The main challenge in obtaining a numerical solution of Eq.~[\ref{renmass}] is that the eigenvalues of the self-consistent boson propagator must be positive, or else the solution is unphysical.
We therefore started from a value of $\lambda$ far from criticality in the disordered phase, which allowed for a positive definite self-consistent boson propagator, and then pushed $\lambda$ towards criticality using solutions at previous values of $\lambda$ as initial conditions.

The allowed $\lambda$ step size to retain positive-definiteness of the boson propagator became progressively smaller as the critical point was approached. Therefore, getting data in the critical regime for the parameter values we analyzed required up to order 10,000 matrix diagonalizations of 25,600$\times$25,600 matrices per disorder realization, as well as the same number of frequency summations over up to order 10,000 Matsubara frequencies per eigenvalue of the matrices and per disorder realization. We executed these operations in a massively parallel manner on a graphics processing unit (GPU), and also utilized multiple GPUs to execute multiple disorder realizations in parallel for high throughput. The use of GPUs reduced matrix diagonalization times down from tens of minutes to just a few seconds, and frequency summation times from tens of seconds to just milliseconds, which made solving the problem possible on a reasonable timescale.

We computed the boson correlator
\begin{align}
D_{ij} (i\Omega) = \frac{1}{M} \left\langle \phi_{ia} (i\Omega) \phi_{ja} (-\Omega) \right\rangle =\sum_{\alpha} \frac{\psi_{\alpha i} \psi_{\alpha j}}{\gamma |\Omega| + \Omega^2/c^2 + e_\alpha}\, \label{Dij}
\end{align}
where $e_\alpha$ and $\psi_{\alpha j}$ are eigenvalues and eigenfunctions of the $\phi$ quadratic form in $\overline{\mathcal{S}}_\phi$, labeled by the index $\alpha = 1 \ldots L^2$ for a $L \times L$ sample. We also computed the spatial Fourier transform
\begin{align}
    D({\bm q}_1, {\bm q}_2, i \Omega) = \frac{1}{L^4}
    \sum_{i,j} D_{ij} (i \Omega) e^{i {\bm q}_1 \cdot {\bm r}_i - i {\bm q}_2 \cdot {\bm r}_j}\,.
\end{align}
where ${\bm r}_j$ is spatial co-ordinate of site $j$, and ${\bm q}_{1,2}$ are two-dimensional momenta.

\begin{figure}
\centering
    \includegraphics[width=\linewidth]{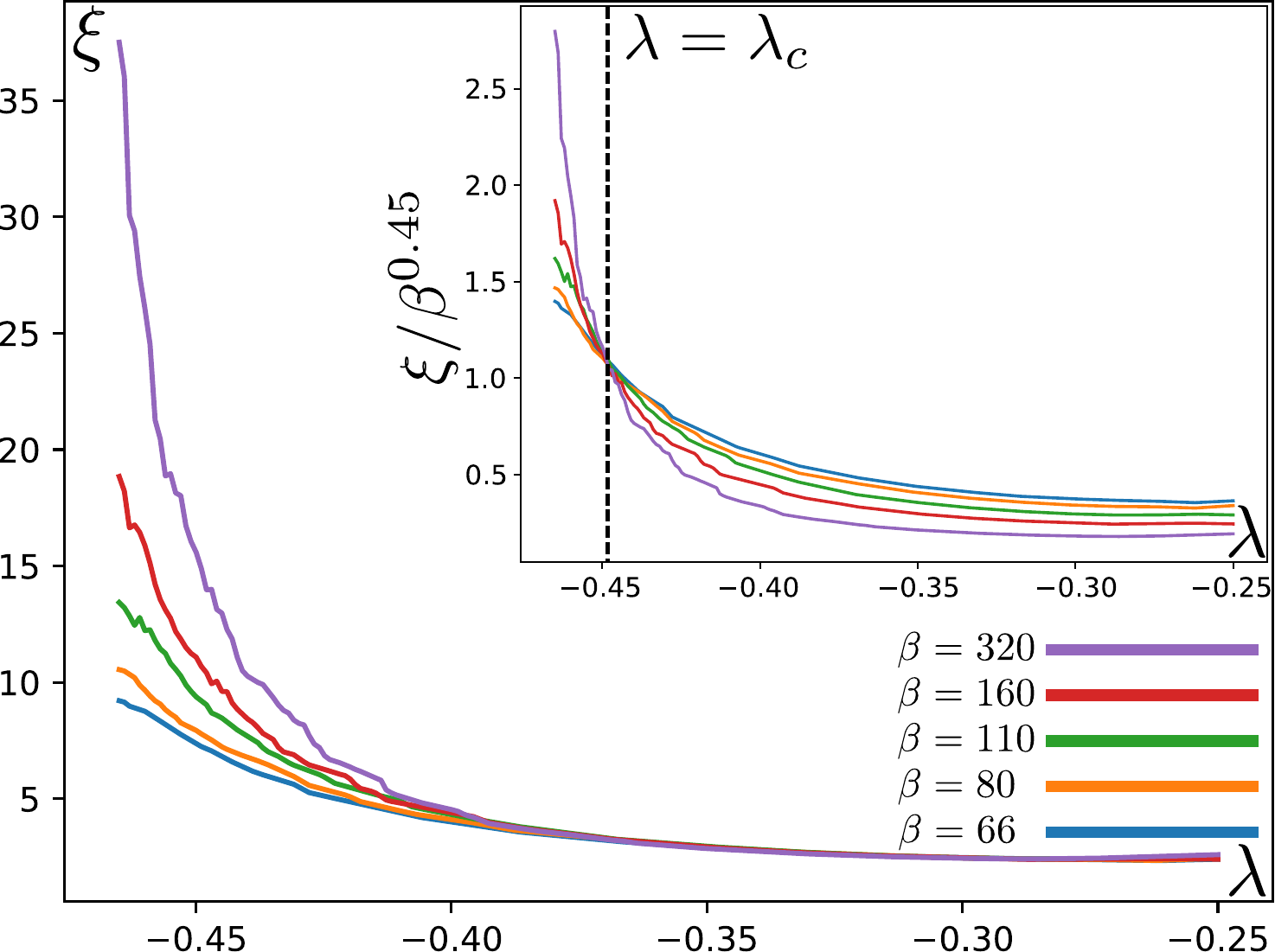}
    \caption{Correlation length as a function of the tuning parameter $\lambda$, for different values of $\beta = 1/T$. (inset) The crossing point where $\xi$ scales as a power law in $\beta$ ($\xi\sim\beta^{0.45}$) estimates the position of the critical point $\lambda_c \approx -0.447$. There is long-range order in $\phi$ for $\lambda < \lambda_c$ at $T=0$.}
    \label{fig:correlation}
\end{figure}

From fitting the decay of the spatially averaged $D_{ij} (i\Omega)$ at large $|{\bm r}_i - {\bm r}_j|$ to $e^{-|{\bm r}_i - {\bm r}_j|/\xi}$, we obtained the correlation length $\xi$, shown in Fig~\ref{fig:correlation}.
A strong $T$ dependence of $\xi$ appears only for smaller values of $\lambda$, and from this we identify the position of the $T = 0$ QCP $\lambda = \lambda_c$ as the point where the correlation length can be best described to be a power law in the inverse temperature $\beta$, {\it i.e.} $\xi\sim \beta^m$. There is long-range order in $\phi$ for $\lambda < \lambda_c$ at $T=0$.

\begin{figure}[ht!]
\centering
    \includegraphics[width=\linewidth]{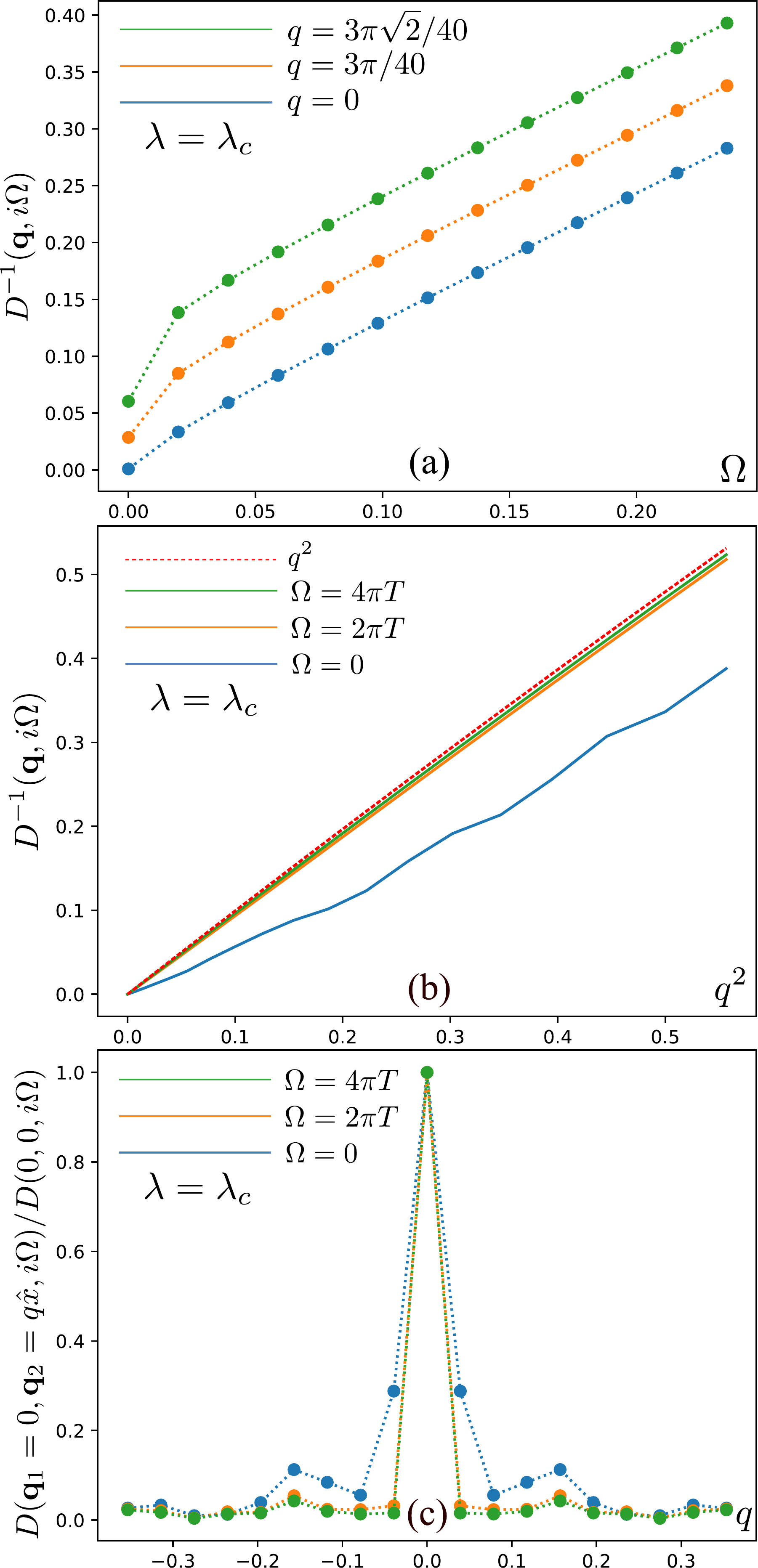}
    \caption{(a) Frequency dependence of $D({\bm q}, i \Omega) \equiv D({\bm q}, {\bm q}, i \Omega)$ at the quantum critical coupling $\lambda=\lambda_c$. Results for other values of $\lambda$ appear in SI Appendix Fig.~\ref{fig:boson_prop_w_supp}. (b) Momentum dependence of $D({\bm q}, i \Omega)$ for different Matsubara frequencies, at the quantum critical coupling $\lambda=\lambda_c$. Results for other values of $\lambda$ appear in SI Appendix Fig.~\ref{fig:boson_prop_supp}. (c) Momentum off-diagonal components of $D({\bm q}_1 = 0, {\bm q}_2, i \Omega)$ for different Matsubara frequencies, again at $\lambda = \lambda_c$. The zeroth Matsubara frequency shows a significant off-diagonal component. These results were obtained for a single disorder sample at $\beta=320$.}
    \label{fig:boson_prop}
\end{figure}

We examine the frequency and momentum dependencies of $D^{-1}({\bm q}, {\bm q}, i \Omega)$ at the QCP $\lambda = \lambda_c$ in Fig~\ref{fig:boson_prop}. We find a $|\Omega|$ frequency dependence and a $q^2$ momentum dependence at all non-zero Matsubara frequencies (Fig.~\ref{fig:boson_prop}a,b), as expected from the averaged theory \cite{Patel:2022gdh}. However, at $\Omega=0$ we find a noisy momentum dependence that is strongly sensitive to the chosen disorder sample, indicating the influence of localized modes at low energies (Fig.~\ref{fig:boson_prop}b). This is confirmed by an examination of $D({\bm q}_1, {\bm q}_2, i\Omega)$ with unequal ${\bm q}_{1,2}$: while the $\Omega \neq 0$ results are strongly peaked at ${\bm q}_1 = {\bm q}$, the $\Omega = 0$ results have large off-diagonal components (Fig.~\ref{fig:boson_prop}c). The frequency dependence also shows a downturn at the zeroth Matsubara frequency, which indicates a change of physics from that of the averaged theory at low energies (Fig.~\ref{fig:boson_prop}a).  

\begin{figure}[h!]
\centering
    \includegraphics[width=\linewidth]{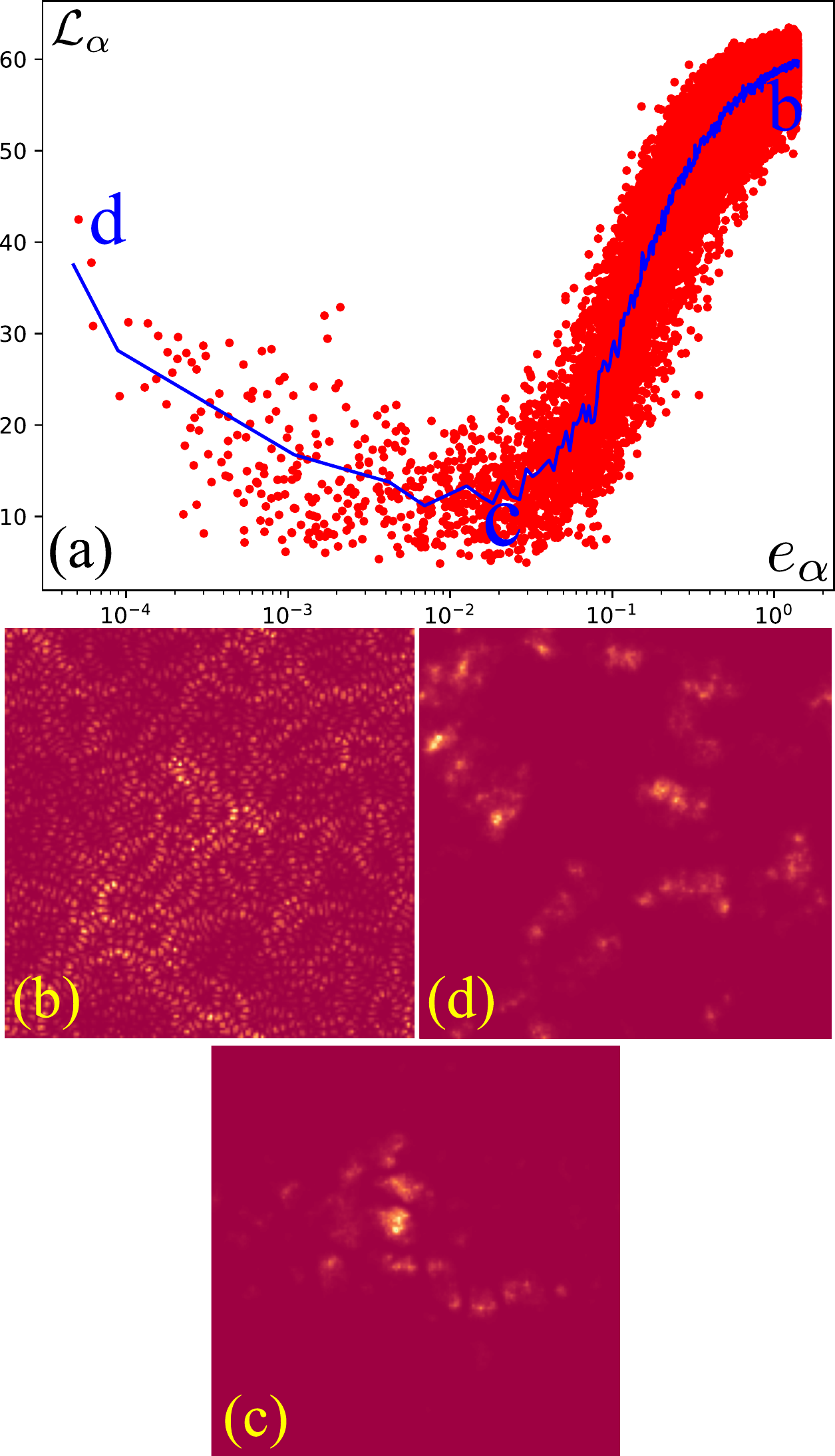}
    \caption{(a) Localization lengths $\mathcal{L}_\alpha$ of the overdamped bosonic modes in a collection of 10 disorder samples as a function of the eigenvalue $e_\alpha$ of the inverse bosonic propagator, at $\lambda = \lambda_c$ and $\beta=800$. The blue curve denotes the average localization length across samples in a given eigenvalue bin. Results for other values of $\lambda$ appear in SI Appendix Fig.~\ref{fig:loc_length_supp}. (b, c, d) Plot of the bosonic eigenstates, $\psi_{\alpha j}$ as a function of ${\bm r}_j$. The corresponding eigenvalue is noted in (a). The delocalized states with higher $e_\alpha$ are superpositions of plane waves, as is apparent from the interference patterns in (b). These transition into strongly localized lower energy states in (c), and start to delocalize again at very low energies in (d). However, the delocalized states in (d) are not wave-like and break translational invariance strongly, as evidenced by the lack of interference patterns.}
    \label{fig:localization}
\end{figure}

A more explicit demonstration of the localization of the low energy modes is presented in Fig.~\ref{fig:localization}.
We compute the localization length by determining the localization volume to be equal to the reciprocal of the inverse participation ratio ($\mathcal{I}_\alpha$) of the normalized low energy eigenvectors $\psi_\alpha$ of $D^{-1}$; the localization length ($\mathcal{L}_\alpha$) is then obtained from the localization volume by assuming an isotropic exponential decay of the eigenvectors:
\beq
\mathcal{I_\alpha} \equiv \sum_j \psi_{\alpha j}^4,~~\mathcal{L_\alpha} \equiv \frac{1}{\sqrt{2\mathcal{I_\alpha}}}.
\eeq
The higher energy modes have a localization length of $\mathcal{O}(L/2)$, as is expected for fully delocalized states in a system with periodic boundary conditions. We expect the universal SYK-type theory of Patel {\it et al.\/} \cite{Patel:2022gdh} to apply at such energies. But at lower energies, Fig.~\ref{fig:localization}a shows a minimum of the localization length, and a slow subsequent increase of the localization length at the lowest energies. This non-monotonic behavior, and the lowest energy increase of the localization length, is just as expected from the physics of the random transverse field Ising model. In the real-space Dasgupta-Ma renormalization group procedure \cite{DasguptaMa}, higher energy localized modes renormalized the couplings of lower energy modes at lower energy, leading to the activated dynamic scaling of the localization length with damping rate $e_\alpha$ \cite{DSF92,DSF95,Motrunich00}
\begin{align}
    \mathcal{L}_\alpha \sim \left[ \ln (1/e_\alpha) \right]^{\mu}\,,
\end{align}
where $\mu$ is an exponent. This logarithmic dependence of length scale on energy is consistent with slow increase of the localization length in Fig.~\ref{fig:localization}a at the lowest energy. There could also be connections of this increasing localization length to `cluster glass' physics \cite{Vlad07}.

\begin{figure}
\centering
    \includegraphics[width=\linewidth]{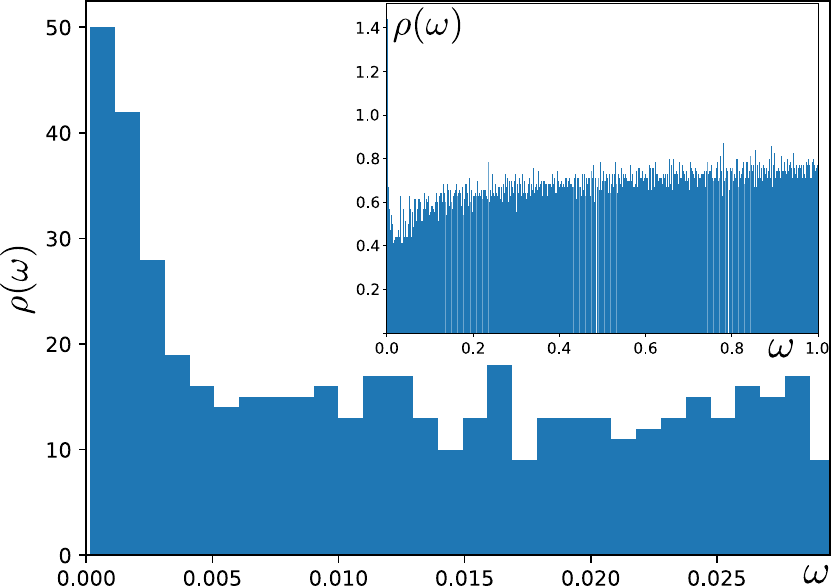}
    \caption{Density of states of boson eigenmodes $\rho(\omega)$ averaged over 10 disorder samples. A larger $\omega$ range is shown in the inset. The localized boson eigenmodes lead to an increase in $\rho(\omega)$ at the lowest energies. These results were obtained at $\beta=320$ and $\lambda=\lambda_c$.}
    \label{fig:DOS}
\end{figure}

We also show a plot of the averaged density of states of eigenmodes of the boson propagator
\begin{align}
   \rho (\omega) = \overline{ \delta (\omega - e_\alpha)}
\end{align}
in Fig.~\ref{fig:DOS}.
The density of states is roughly constant for most of the energy range, as is expected for a $q^2$ boson dispersion in $d=2$, but increases as $\omega\rightarrow0$, where the localized lowest-energy eigenvectors are clustered.

\section{Fermion and transport properties}
\label{sec:fermion}

We now turn to the full model which includes fermionic degrees of freedom. 
The model underlying $\mathcal{S}_\phi + \mathcal{S}_{\phi d} $ in Eq.~[\ref{Sphi}] is that examined by Patel {\it et al.} \cite{Patel:2022gdh}, and involves electrons $c_i$ (we do not write out the electron spin components) coupled to the bosonic modes with imaginary time action
\begin{align}
    \mathcal{S} = &  \mathcal{S}_\phi + \mathcal{S}_c + \mathcal{S}_Y \nonumber \\
    \mathcal{S}_c = & \int d \tau \sum_{{\bm k}} c_{{\bm k}}^\dagger \left( \frac{\partial}{\partial \tau} + \varepsilon_{{\bm k}} \right)  c_{{\bm k}}^{} \nonumber \\
    \mathcal{S}_Y = & \int d\tau \sum_j (g + g^\prime_j )\, \phi_{ja}
    c_{j}^\dagger \sigma^a_j c_j^{}\,. \label{S}
\end{align}
Here ${\bm k}$ is a two-dimensional momentum, $\varepsilon_{\bm k}$ is the electron dispersion with a simple convex Fermi surface, $\sigma^a_j$ is a fixed coupling matrix depending upon the nature of the field $\phi$, and the Yukawa coupling $g + g_j^\prime$ has a spatially random component obeying
\begin{align}
    \overline {g_j^\prime} = 0 \quad, \quad  \overline {g_j^\prime g_k^\prime} = \delta_{jk} g^{\prime 2}\,.
\end{align}
Patel {\it et al.} \cite{Patel:2022gdh} argued that the spatial randomness in $\lambda + \lambda_j^\prime$ could be `gauged away' by rescaling $\phi_j$, and then analyzed $\mathcal{S}$ with a spatially independent $\lambda$, averaging over the disorder $g_j'$ along the lines of the Yukawa-SYK model \cite{Wang:2019bpd,Esterlis:2019ola}. We expect that this procedure should be applicable as long as we are in the regime with extended bosonic eigenmodes, above the minimum in Fig.~\ref{fig:localization}a. But we do not expect it to be applicable in the strong disorder regime associated with the localized bosonic eigenmodes below the minimum in Fig.~\ref{fig:localization}a. We also note a perturbative study of localization effects \cite{Foster_22} on a model closely related to Eq.~[\ref{S}] which strong corrections.

\begin{figure}
\centering
    \includegraphics[width=\linewidth]{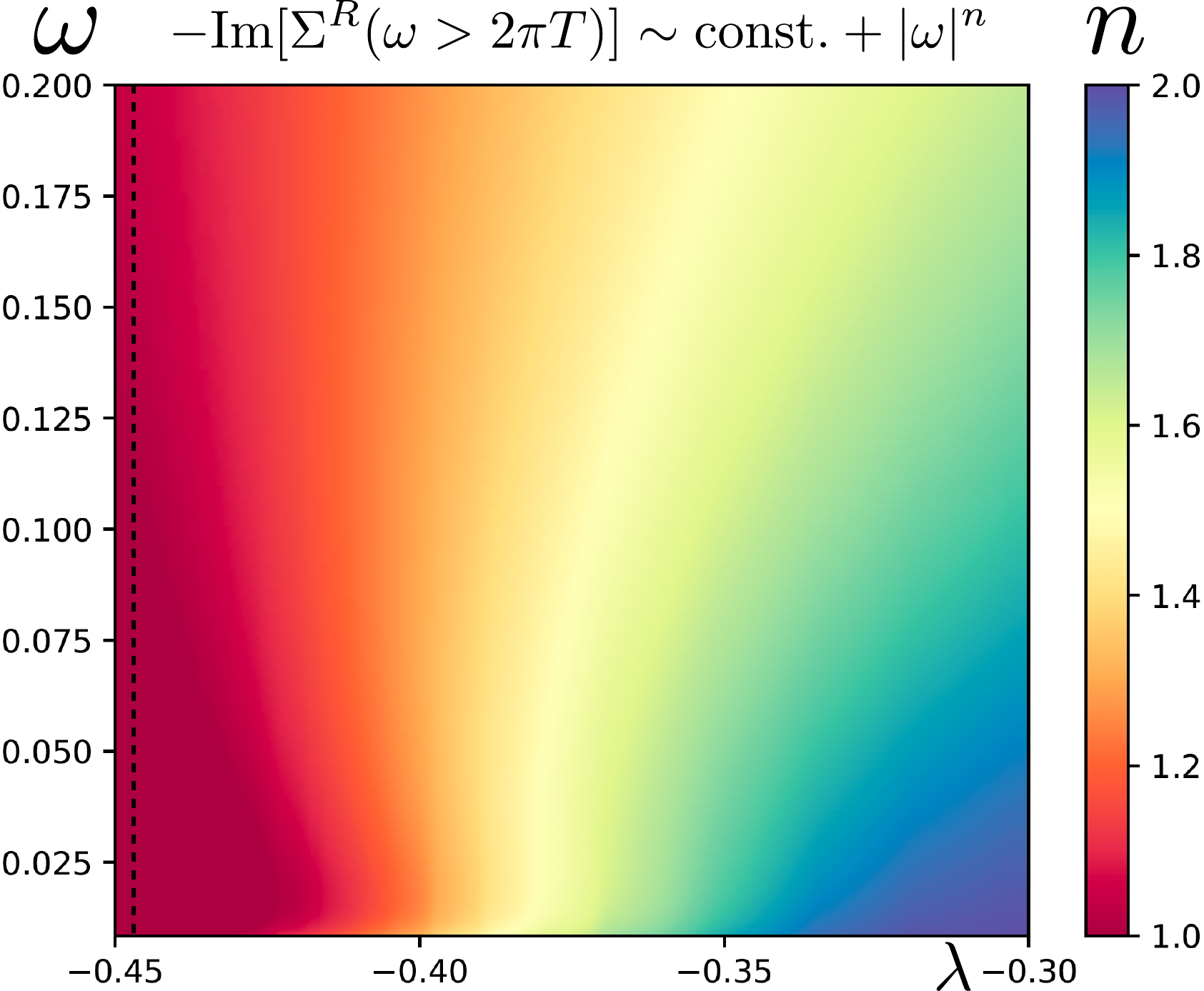}
    \caption{Power-law scaling in the frequency ($\omega \gg T$) dependence of the transport scattering rate inferred from the contribution to the single particle self energy from spatially random ($g'$) interactions. The dashed line represents $\lambda=\lambda_c$. These results were obtained at $\beta=800$ and are an average over $10$ disorder samples. An exponent $n\sim1$ is seen over an extended region away from the QCP that becomes wider in $\lambda$ at lower energy scales, thereby resembling the results of Ref.~\cite{Hussey_foot}. The exponent $n$ is determined using $n = -d\ln(\mathrm{Im}[\Sigma^R(\omega)]-\mathrm{Im}[\Sigma^R(0)])/d\ln\omega$.}
    \label{fig:scattering}
\end{figure}

Here, we wish to describe the consequences of the crossover in the bosonic eigenmodes in Fig.~\ref{fig:localization} in the electronic spectrum. To this end, we will use the bosonic eigenmodes of Section~\ref{sec:numbos} to compute the electron self energy perturbatively in $g+g_j'$, assuming that the electronic eigenmodes remain extended. For the extended bosonic eigenmodes, it has been argued \cite{Patel:2022gdh} that the fermion self energy due to the spatially uniform coupling $g$ cancels in the computation of transport properties. For the localized bosonic modes, the influence of $g$ and $g_j'$ in the electronic self energy should be similar, as the randomness in the eigenfunctions $\psi_{\alpha j}$ ensures lack of momentum conservation. So for transport properties, it is adequate to follow the simpler procedure of computing the electronic self energy only from $g_j'$, and using the imaginary part of the retarded self energy as a proxy for the transport scattering rate. We do not self-consistently recompute the Landau damping for the boson $\phi$ (as was done in earlier work \cite{Ilya1}), as the fermions remain extended, and we do not expect significant spatial dependence. We therefore compute the average perturbative electronic self energy via
\begin{align}
    \Sigma (i \omega) & = - i \pi g^{\prime 2} \mathcal{N}_0 \frac{T}{L^2} \sum_{j,\Omega} \mbox{sgn} (\omega + \Omega) D_{jj} (i \Omega) \nn
                      & = - i \pi g^{\prime 2} \mathcal{N}_0 \frac{T}{L^2} \sum_{\alpha,\Omega}\frac{\mbox{sgn} (\omega + \Omega)}{\gamma   |\Omega| + \Omega^2/c^2 + e_\alpha},
    \label{Sigmai}
\end{align}
where $\mathcal{N}_0$ is the density of electronic states at the Fermi level, associated with the dispersion $\varepsilon_{\bm k}$. The last expression is only valid for the large $M$ self-consistent approach, and
now the Matsubara summation can be performed exactly in closed form (see SI Appendix Eq.~[\ref{analytical_self_en}]), with the sum over eigenvalues $e_\alpha$ subsequently performed numerically. Therefore, an important advantage of this computational procedure is that we can perform an exact analytic continuation to real frequencies, $i\omega\rightarrow\omega+i0^+$, and then obtain the retarded fermion self energy $\Sigma^R(\omega)$ on the real frequency axis. Taking the imaginary part of $\Sigma^R(\omega)$, we obtain the results for the $\omega$ dependence of the transport scattering rate shown in Fig.~\ref{fig:scattering}. We find a power-law dependence on $\omega$ for $\omega\gg T$. The exponent $n$ of the power law is approximately $1$ for a range of $\lambda > \lambda_c$, indicating the extension of the strange metal (which is defined by an exponent $n = 1$ in the transport scattering rate) into a `quantum critical phase' away from the QCP.

\begin{figure}[ht!]
\centering
    \includegraphics[width=\linewidth]{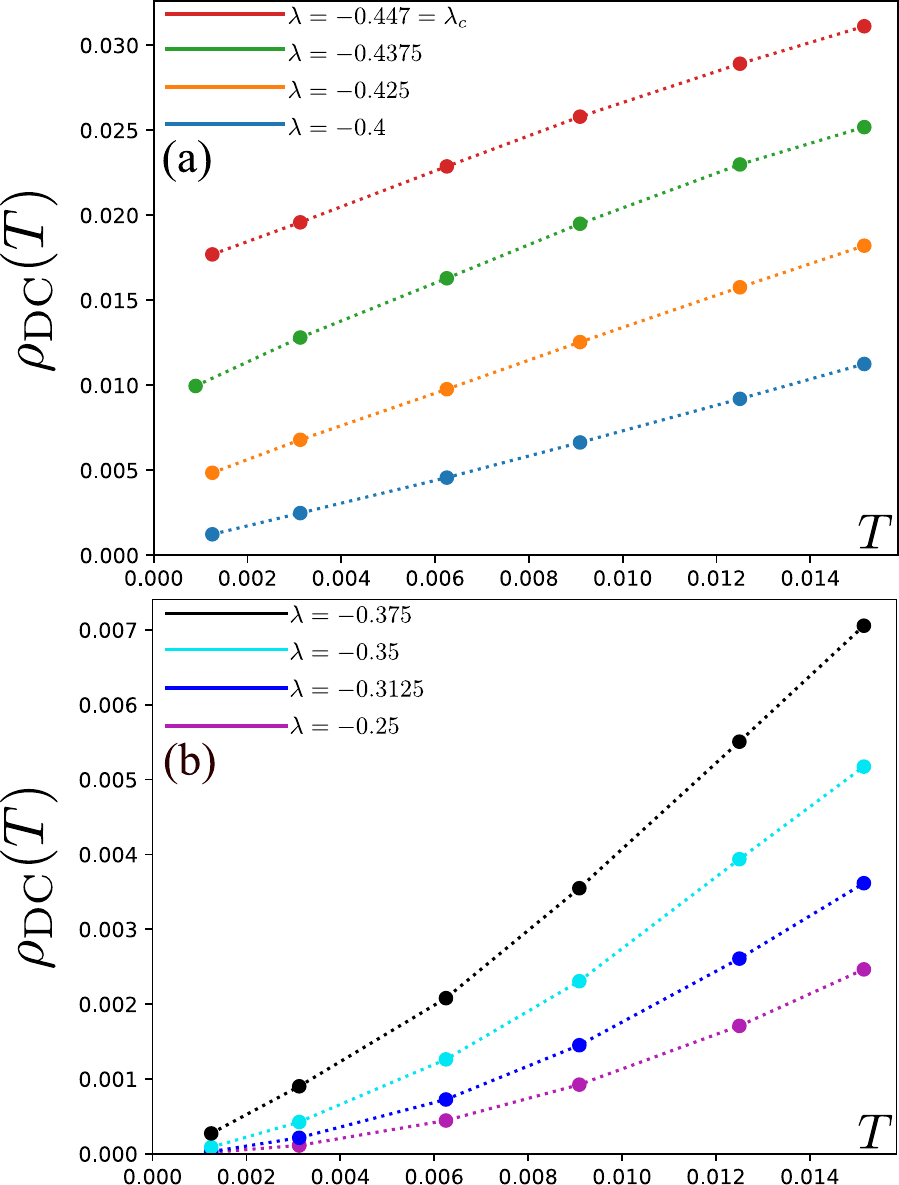}
    \caption{(a) DC resistivity $\rho_\mathrm{DC}(T)$ vs temperature $T$. A roughly linear dependence on $T$ is seen over an extended region away from the QCP, along with a residual resistivity, thereby resembling the results of Ref.~\cite{Hussey_foot}. (b) DC resistivity further away from the QCP, showing superlinear scaling in $T$, and eventually transitioning into the Fermi-liquid $T^2$ scaling far away from the QCP. These results were obtained by averaging over 10 disorder samples.}
    \label{fig:rho_DC}
\end{figure}

Finally, we compute the DC resistivity $\rho_{\mathrm{DC}}$ from $\mathrm{Im}[\Sigma^R(\omega)]$, using the standard relation \cite{Patel:2018mag} that is valid for transport determined by the non-momentum conserving scattering arising from the spatially random part $g'$ of the Yukawa coupling:
\begin{equation}
\frac{1}{\rho_\mathrm{DC}(T)} = \frac{\pi v_F^2 \mathcal{N}_0}{8T}\int_{-\infty}^{\infty}\frac{d\omega}{2\pi}\mathrm{sech}^2\left(\frac{\omega}{2T}\right)\frac{1}{\mathrm{Im}[\Sigma^R(\omega)]},
\label{Drude}
\end{equation}
where $v_F$ is the average Fermi velocity of the electrons. We plot the $T$-dependence of the DC resistivity in Fig.~\ref{fig:rho_DC}. A linear temperature dependence of the resistivity is seen for a significant range of $\lambda > \lambda_c$, again indicating a `quantum critical phase'. Eventually, for $\lambda \gg \lambda_c$, the temperature dependence crosses over to the quadratic scaling expected in a Fermi liquid. Interestingly, we also observe a finite residual resistivity, that becomes significant as $\lambda\rightarrow\lambda_c$. Its origin can be traced back to the large boson density of states at $\omega\rightarrow0$ shown in Fig.~\ref{fig:DOS}. From Eq.~[\ref{Sigmai}], a cluster of near-zero eigenvalues $e_\alpha$ can be seen to produce a nearly $\omega$-independent offset in $\Sigma(i\omega)$ and $\mathrm{Im}[\Sigma^R(\omega)]$, which translates into a residual resistivity through Eq.~[\ref{Drude}]. The physical interpretation of this effect is simple - the lowest energy boson eigenstates are localized in nature and are also nearly frozen with very slow dynamics, and therefore simply act as local elastic impurity scatterers of the electrons, giving rise to a residual resistivity.

\section{$M=1$ Ising scalar}
\label{sec:scalar}

When $M = 1$, the large $M$ saddle point Eqs. (\ref{renmass}) are no longer applicable. We therefore simulate the theory of Eq.~[\ref{Sphi}] using a Hybrid Monte Carlo (HMC) algorithm. We use a HMC method recently developed for critical fermionic theories \cite{LuntsHMC}, but without the fermions, running on multiple GPUs to sample over many disorder configurations. We find that the sharp QCP becomes smeared over some region of $\lambda$, as indicated by the correlation length and the susceptibility and shown in Fig. \ref{fig:xi, B_c and chi M = 1}. 
\begin{figure}[ht!]
\centering
    \includegraphics[width=\linewidth]{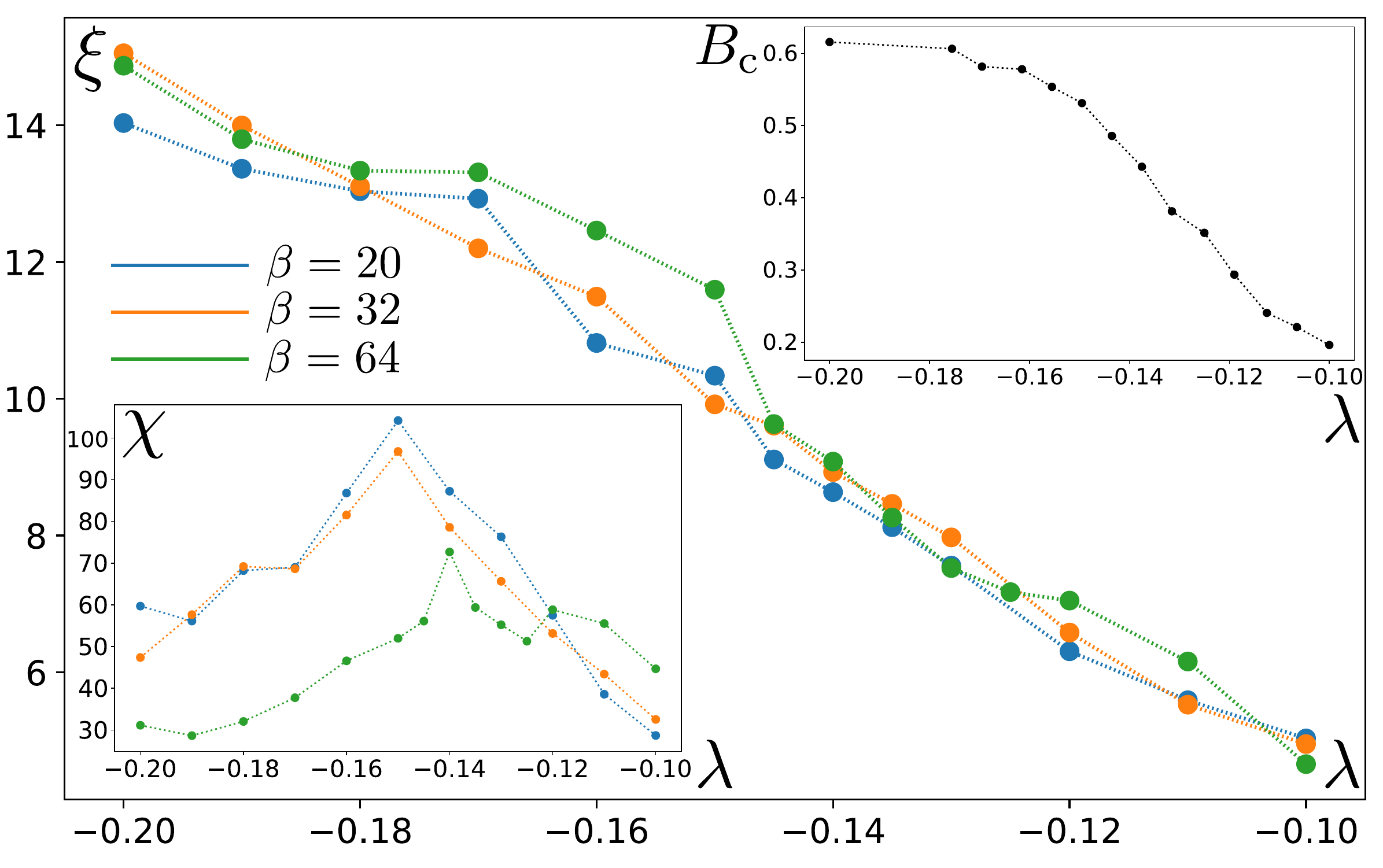}
    \caption{Observables describing the rounded phase transition for the $M = 1$ Ising scalar. The main figure shows the correlation length of the equal-time propagator, $\xi$, plotted as a function of $\lambda$, for an $L = 80$ system and with $40$ disorder configurations. Unlike the $M \geq 2$ case, $\xi$ does not diverge (or even change) with temperature in the entire critical region, indicating the absence of a sharp transition. The inset in the upper right corner shows the Binder cumulant, $B_c = 1 - \overline{\langle \Phi^4 \rangle/(3 \langle \Phi^2 \rangle^2)}$, where $\Phi \equiv \phi(\Omega = 0, \bm{q} = 0)$, plotted as a function of $\lambda$, for $L = 64, \beta = 32$ and with $120$ disorder configurations. The shape of the curve tells us the approximate `smeared critical' region, as well as the $\lambda$ beyond which $B_c$ is close to the limiting value of $0.66$ and can be considered `ordered'. We plot only one value of $\beta, L$ as no crossing point will occur in this case. The inset in the lower left corner shows the uniform static susceptibility $\chi$. All curves are for $L = 80$, and the $\beta = 20,32$ ($\beta = 64$) curves use $100-200$ ($40$) disorder configurations. As expected, the susceptibility peak does not become sharper with lower $T$, as is the case for QCPs. However, the presence of a peak indicates a smaller region of $\lambda$ where we expect the most critical-like behavior.}
    \label{fig:xi, B_c and chi M = 1}
\end{figure}
This is consistent with the predictions of Refs. \cite{VojtaIsing1,VojtaIsing2}. In this `smeared critical' region, the disorder variance is significantly enhanced compared to the $M \geq 2$ case, with hundreds of disorder configurations required to smooth out many of the observables for which only a few configurations were sufficient at large $M$. The wavefunction localization lengths $\mathcal{L}_\alpha$ are shown in SI Appendix 
Fig.~\ref{fig:LLs_supp} and behave largely the same as in Fig. \ref{fig:localization}a, but with a few differences: there are significantly fewer localized eigenmodes, and the de-localized but translation-symmetry breaking lowest eigenmodes get spectrally separated from the localized ones as $\lambda$ is lowered. The former is again consistent with the absence of a Griffiths phase where a large density of localized eigenmodes give a critical spectral density at low energies. The latter is a novel observation, and we attribute these states to ordered puddles which are no longer fluctuating do to the discrete symmetry breaking. More plots of bosonic properties are shown in the SI Appendix. 

\begin{figure}[ht!]
\centering
    \includegraphics[width=\linewidth]{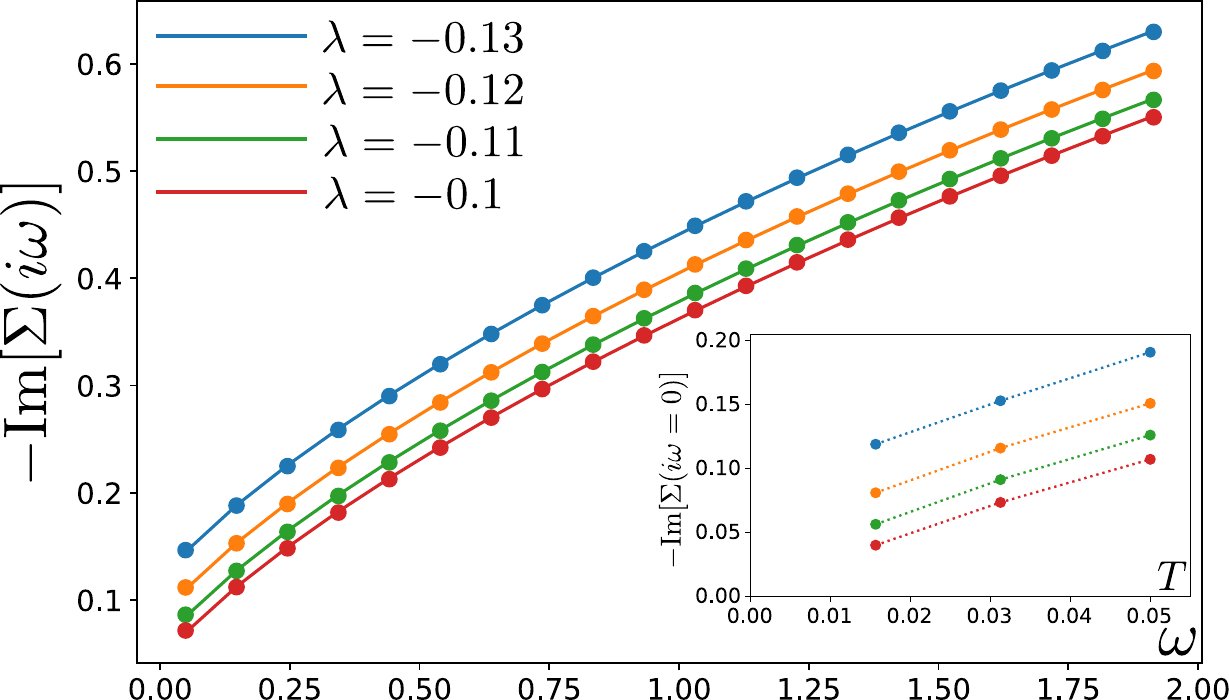}
    \caption{The fermion self-energy $-\mathrm{Im}[\Sigma(i \omega)]$ as a function of Matsubara frequency $\omega$, in the `smeared critical' region for $M=1$. All curves are for an $L = 80$ system at $\beta = 64$, using an average over $100$-$200$ disorder realizations. The lines are fits to the MFL form $a \, \omega \log(b/|\omega|) + c$, which has a UV cutoff $b$ and a $\omega=0$ scattering rate $c$. The MFL form is very well fit for all $\lambda$ values shown, with coefficients for all curves in the ranges: $0.083 < a < 0.088$, $41 < b < 46$, and $0.04 < c < 0.12$ (increasing with decreasing $\lambda$). (inset) The extrapolated value of $-\mathrm{Im}[\Sigma(i\omega=0)]$ from the MFL fit vs temperature $T$, showing a $T$ dependence of roughly $-T\ln T$ along with a residual scattering rate, as expected for a MFL.}
    \label{fig:Sigma vs omega M = 1}
\end{figure}

Although the HMC solution does not enable us to analytically continue the fermion self-energy, we can still evaluate it on the Matsubara axis using the first expression of Eq.~[\ref{Sigmai}]. In the entire `smeared critical' region, we observe a very distinct marginal Fermi-liquid (MFL) scaling of $\mathrm{Im}[\Sigma(i \omega)]$, as shown in Fig. \ref{fig:Sigma vs omega M = 1}.
Based on the arguments of Patel {\it et al.} \cite{Patel:2022gdh} and those in the previous section, upon analytic continuation to real frequencies this MFL self energy is what leads to strange metal behavior with linear-in-temperature and linear-in-frequency transport scattering rates.

\section{Discussion}

The recent universal theory of strange metals by Patel {\it et al.\/} \cite{Patel:2022gdh} considered the action $\mathcal{S}$ in Eq.~[\ref{S}] of electrons $c_j$ coupled to quantum critical bosonic scalars $\phi_{ja}$. They argued that the random `mass' spatial disorder $\lambda_j^\prime$ in Eq.~[\ref{Sphi}] could be absorbed by a rescaling of the scalar fields $\phi_{ja}$, resulting in an enhancement of the spatial disorder in the Yukawa coupling $g_j^\prime$ in Eq.~[\ref{S}].
They performed a self-consistent and self-averaging analysis of the resulting action, similar to that required for the exact solution of the SYK model \cite{Chowdhury:2021qpy}.  This theory matched numerous observations in strange metals, including the $T$-linear resistivity, the Planckian relaxation time, the $T \ln(1/T)$ specific heat, and the $1/\omega$ optical conductivity.

The present paper has focused closer attention on the role of the random mass spatial disorder $\lambda_j^\prime$. We have shown that the rescaling procedure of Patel~{\it et al.} \cite{Patel:2022gdh} remains valid in an intermediate temperature regime where the eigenmodes of the zero frequency boson propagator remain extended. However, new physics emerges at low temperatures when the boson eigenmodes localize, resulting in an extended regime of strange metal behavior away from the QCP. This extended regime is proposed as an explanation of observations by Cooper {\it et al.} \cite{Hussey_foot} and Greene {\it et al.} \cite{Greene_rev}.

Our key results for the localization of the boson eigenmodes for $M \geq 2$ appear in Fig.~\ref{fig:localization}. At higher energies, the bosonic eigenmodes are extended, as in Fig.~\ref{fig:localization}b. The extended bosonic eigenmodes have a density of states which is independent of energy, as shown in Fig.~\ref{fig:DOS}, and as found in the SYK-type analysis by Patel {\it et al.\/} \cite{Patel:2022gdh}. This constant density of states results in a linear-in-$T$ resistivity (Fig.~\ref{fig:rho_DC}), that we found to extend away from QCP. Fig.~\ref{fig:localization}a shows a minimum in the localization length below which the localization length shows a logarithmically slow increase with decreasing energy. We have argued that this low energy regime is described \cite{Hoyos07,Hoyos09} by the strong-disorder Griffiths regime of the random transverse field Ising model \cite{DSF92,DSF95}. We computed the effect of these localized eigenmodes on electronic transport perturbatively, and showed that they produce a significant contribution the residual resistivity as the QCP is approached. However, it would be worthwhile to examine the contributions of the localized modes more completely in future work.

For the case of the $M = 1$ Ising scalar, the localized modes are far fewer in number, which leads to an absence of a Griffiths phase and a `smeared critical' region which replaces the sharp QCP of $M \geq 2$. Like in the $M\geq2$ case, this region also shows strange metal behavior at finite energies over an extended range of the critical tuning parameter. However, due to the finite correlation length, we expect the strange metal behavior to not extend all the way down to zero temperature, and instead give way to Fermi liquid behavior at the lowest energy scales, unlike in the $M\geq2$ case. This is of direct relevance to experiments on strange metals near possible Ising-nematic QCPs such as those in FeSe studied in Ref. \cite{Hussey_FeSe}. Ref. \cite{Hussey_FeSe} suggests that the low temperature strange metal behavior observed in FeSe might be due to antiferromagnetic ($M=3$) fluctuations, rather than Ising-nematic ($M=1$) fluctuations, which would be in alignment with our conclusions about a Fermi liquid ground state for $M=1$. 

It would also be interesting to directly observe the dynamics of the localized overdamped eigenmodes in strange metals.  These eigenmodes resemble `two-level systems' in glasses, and perhaps similar experimental methods can be used \cite{Bashan23,Wolynes-rev,Gruebele19}, or those used to image nanoscale electron flow \cite{Yacoby20}. Additionally, it might be possible to see indirect signatures of these modes in low energy dynamical structure factors $S^R(\mathbf{q}, \omega)$ (which could be in either spin or charge channels depending upon the physical origin of the bosonic modes). We would expect, for instance, the wavevector-integrated structure factor $\mathrm{Im}[S^R(\omega)] \equiv \int d^2\mathbf{q}~\mathrm{Im}[S^R(\mathbf{q},\omega)]$ to show an upturn like in Fig.~\ref{fig:DOS} starting at around $\omega \sim 0.1 - 1$ meV, going by the energy scales in Ref.~\cite{Hussey_foot}. Such upturns should also occur in structure factors $S^R(\mathbf{q}, \omega)$ measured at a fixed wavevector $\mathbf{q}$.

\subsection*{Acknowledgements}

We thank Erez Berg, Hitesh Changlani, Sankar Das Sarma, Adrian Del Maestro, Yu He, Steve Kivelson, Akshat Pandey, Sri Raghu, J\"org Schmalian, and T.~Senthil for valuable discussions.
This research was supported by the  U.S. National Science Foundation grant No.
DMR-2245246, the Harvard Quantum Initiative Postdoctoral Fellowship in Science and Engineering, and by the Simons Collaboration on Ultra-Quantum Matter which is a grant from the Simons Foundation (651440, S.S.). The Flatiron Institute is a division of the Simons Foundation.  This research was supported in part by grant NSF PHY-1748958 to the Kavli Institute for Theoretical Physics (KITP).

\bibliography{fermi}

\providecommand{\href}[2]{#2}\begingroup\raggedright\begin{thebibliography}{10}

\bibitem{Hartnoll:2021ydi}
S.A.~Hartnoll and A.P.~Mackenzie, \emph{{Colloquium: Planckian dissipation in
  metals}}, \href{https://doi.org/10.1103/RevModPhys.94.041002}{\emph{Rev. Mod.
  Phys.} {\bfseries 94} (2022) 041002}
  [\href{https://arxiv.org/abs/2107.07802}{{\ttfamily 2107.07802}}].

\bibitem{Michon22}
B.~{Michon}, C.~{Berthod}, C.W.~{Rischau}, A.~{Ataei}, L.~{Chen}, S.~{Komiya}
  et~al., \emph{{Reconciling scaling of the optical conductivity of cuprate
  superconductors with Planckian resistivity and specific heat}},
  \href{https://doi.org/10.1038/s41467-023-38762-5}{\emph{Nature
  Communications} {\bfseries 14} (2023) 3033}
  [\href{https://arxiv.org/abs/2205.04030}{{\ttfamily 2205.04030}}].

\bibitem{Patel:2022gdh}
A.A.~Patel, H.~Guo, I.~Esterlis and S.~Sachdev, \emph{{Universal theory of
  strange metals from spatially random interactions}},
  \href{https://doi.org/10.1126/science.abq6011}{\emph{Science} {\bfseries 381}
  (2023) abq6011} [\href{https://arxiv.org/abs/2203.04990}{{\ttfamily
  2203.04990}}].

\bibitem{VojtaRMP}
H.V.~{L{\"o}hneysen}, A.~{Rosch}, M.~{Vojta} and P.~{W{\"o}lfle},
  \emph{{Fermi-liquid instabilities at magnetic quantum phase transitions}},
  \href{https://doi.org/10.1103/RevModPhys.79.1015}{\emph{Rev. Mod. Phys.}
  {\bfseries 79} (2007) 1015}
  [\href{https://arxiv.org/abs/cond-mat/0606317}{{\ttfamily
  cond-mat/0606317}}].

\bibitem{Chowdhury:2021qpy}
D.~Chowdhury, A.~Georges, O.~Parcollet and S.~Sachdev, \emph{{Sachdev-Ye-Kitaev
  models and beyond: Window into non-Fermi liquids}},
  \href{https://doi.org/10.1103/RevModPhys.94.035004}{\emph{Rev. Mod. Phys.}
  {\bfseries 94} (2022) 035004}
  [\href{https://arxiv.org/abs/2109.05037}{{\ttfamily 2109.05037}}].

\bibitem{Hussey_foot}
R.A.~Cooper, Y.~Wang, B.~Vignolle, O.J.~Lipscombe, S.M.~Hayden, Y.~Tanabe
  et~al., \emph{{Anomalous Criticality in the Electrical Resistivity of
  La$_{2-x}$Sr$_x$CuO$_4$}},
  \href{https://doi.org/10.1126/science.1165015}{\emph{Science} {\bfseries 323}
  (2009) 603}.

\bibitem{Greene_rev}
R.L.~{Greene}, P.R.~{Mandal}, N.R.~{Poniatowski} and T.~{Sarkar}, \emph{{The
  Strange Metal State of the Electron-Doped Cuprates}},
  \href{https://doi.org/10.1146/annurev-conmatphys-031119-050558}{\emph{Annual
  Review of Condensed Matter Physics} {\bfseries 11} (2020) 213}
  [\href{https://arxiv.org/abs/1905.04998}{{\ttfamily 1905.04998}}].

\bibitem{Bashan23}
N.~{Bashan}, E.~{Tulipman}, J.~{Schmalian} and E.~{Berg}, \emph{{Tunable
  non-Fermi liquid phase from coupling to two-level systems}},
  \href{https://doi.org/10.48550/arXiv.2310.07768}{\emph{arXiv e-prints} (2023)
  arXiv:2310.07768} [\href{https://arxiv.org/abs/2310.07768}{{\ttfamily
  2310.07768}}].

\bibitem{Milica89}
M.~Milovanovi\ifmmode~\acute{c}\else \'{c}\fi{}, S.~Sachdev and R.N.~Bhatt,
  \emph{Effective-field theory of local-moment formation in disordered metals},
  \href{https://doi.org/10.1103/PhysRevLett.63.82}{\emph{Phys. Rev. Lett.}
  {\bfseries 63} (1989) 82}.

\bibitem{BhattFisher}
R.N.~Bhatt and D.S.~Fisher, \emph{Absence of spin diffusion in most random
  lattices}, \href{https://doi.org/10.1103/PhysRevLett.68.3072}{\emph{Phys.
  Rev. Lett.} {\bfseries 68} (1992) 3072}.

\bibitem{Vlad05a}
V.~{Dobrosavljevi{\'c}} and E.~{Miranda}, \emph{{Absence of Conventional
  Quantum Phase Transitions in Itinerant Systems with Disorder}},
  \href{https://doi.org/10.1103/PhysRevLett.94.187203}{\emph{Phys. Rev. Lett.}
  {\bfseries 94} (2005) 187203}
  [\href{https://arxiv.org/abs/cond-mat/0408336}{{\ttfamily
  cond-mat/0408336}}].

\bibitem{Vlad05b}
D.~{Tanaskovi{\'c}}, V.~{Dobrosavljevi{\'c}} and E.~{Miranda},
  \emph{{Spin-Liquid Behavior in Electronic Griffiths Phases}},
  \href{https://doi.org/10.1103/PhysRevLett.95.167204}{\emph{Phys. Rev. Lett.}
  {\bfseries 95} (2005) 167204}
  [\href{https://arxiv.org/abs/cond-mat/0412100}{{\ttfamily
  cond-mat/0412100}}].

\bibitem{Vlad07}
M.J.~{Case} and V.~{Dobrosavljevi{\'c}}, \emph{{Quantum Critical Behavior of
  the Cluster Glass Phase}},
  \href{https://doi.org/10.1103/PhysRevLett.99.147204}{\emph{Phys. Rev. Lett.}
  {\bfseries 99} (2007) 147204}
  [\href{https://arxiv.org/abs/cond-mat/0612127}{{\ttfamily
  cond-mat/0612127}}].

\bibitem{Vojta_rev}
T.~{Vojta}, \emph{{Phases and phase transitions in disordered quantum
  systems}},  in \emph{Lectures on the Physics of Strongly Correlated Systems
  XVII: Seventeenth Training Course in the Physics of Strongly Correlated
  Systems}, A.~{Avella} and F.~{Mancini}, eds., vol.~1550 of \emph{American
  Institute of Physics Conference Series}, pp.~188--247, Aug., 2013,
  \href{https://doi.org/10.1063/1.4818403}{DOI}
  [\href{https://arxiv.org/abs/1301.7746}{{\ttfamily 1301.7746}}].

\bibitem{DCSS15b}
D.~{Chowdhury} and S.~{Sachdev}, \emph{{Higgs criticality in a two-dimensional
  metal}}, \href{https://doi.org/10.1103/PhysRevB.91.115123}{\emph{Phys. Rev.
  B} {\bfseries 91} (2015) 115123}
  [\href{https://arxiv.org/abs/1412.1086}{{\ttfamily 1412.1086}}].

\bibitem{Aldape22}
E.E.~{Aldape}, T.~{Cookmeyer}, A.A.~{Patel} and E.~{Altman}, \emph{{Solvable
  theory of a strange metal at the breakdown of a heavy Fermi liquid}},
  \href{https://doi.org/10.1103/PhysRevB.105.235111}{\emph{Phys. Rev. B}
  {\bfseries 105} (2022) 235111}
  [\href{https://arxiv.org/abs/2012.00763}{{\ttfamily 2012.00763}}].

\bibitem{Sachdev:2023fim}
S.~{Sachdev}, \emph{{Strange metals and black holes: insights from the
  Sachdev-Ye-Kitaev model}},
  \href{https://doi.org/10.1093/acrefore/9780190871994.013.48}{\emph{Oxford
  Research Encyclopedia in Physics} (2023) }
  [\href{https://arxiv.org/abs/2305.01001}{{\ttfamily 2305.01001}}].

\bibitem{Hoyos07}
J.A.~{Hoyos}, C.~{Kotabage} and T.~{Vojta}, \emph{{Effects of Dissipation on a
  Quantum Critical Point with Disorder}},
  \href{https://doi.org/10.1103/PhysRevLett.99.230601}{\emph{Phys. Rev. Lett.}
  {\bfseries 99} (2007) 230601}
  [\href{https://arxiv.org/abs/0705.1865}{{\ttfamily 0705.1865}}].

\bibitem{Hoyos09}
T.~{Vojta}, C.~{Kotabage} and J.A.~{Hoyos}, \emph{{Infinite-randomness quantum
  critical points induced by dissipation}},
  \href{https://doi.org/10.1103/PhysRevB.79.024401}{\emph{Phys. Rev. B}
  {\bfseries 79} (2009) 024401}
  [\href{https://arxiv.org/abs/0809.2699}{{\ttfamily 0809.2699}}].

\bibitem{CastroNeto1}
A.H.~Castro~Neto and B.A.~Jones, \emph{Non-fermi-liquid behavior in u and ce
  alloys: Criticality, disorder, dissipation, and griffiths-mccoy
  singularities}, \href{https://doi.org/10.1103/PhysRevB.62.14975}{\emph{Phys.
  Rev. B} {\bfseries 62} (2000) 14975}.

\bibitem{Schmalian02}
A.J.~{Millis}, D.K.~{Morr} and J.~{Schmalian}, \emph{{Quantum Griffiths effects
  in metallic systems}},
  \href{https://doi.org/10.1103/PhysRevB.66.174433}{\emph{Phys. Rev. B}
  {\bfseries 66} (2002) 174433}
  [\href{https://arxiv.org/abs/cond-mat/0208396}{{\ttfamily
  cond-mat/0208396}}].

\bibitem{Schmalian05}
T.~{Vojta} and J.~{Schmalian}, \emph{{Quantum Griffiths effects in itinerant
  Heisenberg magnets}},
  \href{https://doi.org/10.1103/PhysRevB.72.045438}{\emph{Phys. Rev. B}
  {\bfseries 72} (2005) 045438}
  [\href{https://arxiv.org/abs/cond-mat/0405609}{{\ttfamily
  cond-mat/0405609}}].

\bibitem{DSF92}
D.S.~Fisher, \emph{{Random transverse field Ising spin chains}},
  \href{https://doi.org/10.1103/PhysRevLett.69.534}{\emph{Phys. Rev. Lett.}
  {\bfseries 69} (1992) 534}.

\bibitem{DSF95}
D.S.~Fisher, \emph{{Critical behavior of random transverse-field Ising spin
  chains}}, \href{https://doi.org/10.1103/PhysRevB.51.6411}{\emph{Phys. Rev. B}
  {\bfseries 51} (1995) 6411}.

\bibitem{Motrunich00}
O.~{Motrunich}, S.-C.~{Mau}, D.A.~{Huse} and D.S.~{Fisher},
  \emph{{Infinite-randomness quantum Ising critical fixed points}},
  \href{https://doi.org/10.1103/PhysRevB.61.1160}{\emph{Phys. Rev. B}
  {\bfseries 61} (2000) 1160}
  [\href{https://arxiv.org/abs/cond-mat/9906322}{{\ttfamily
  cond-mat/9906322}}].

\bibitem{Pandey23}
A.~{Pandey}, A.~{Mahadevan} and A.~{Cowsik}, \emph{{Random geometry at an
  infinite-randomness fixed point}},
  \href{https://doi.org/10.1103/PhysRevB.108.064201}{\emph{Phys. Rev. B}
  {\bfseries 108} (2023) 064201}
  [\href{https://arxiv.org/abs/2304.10564}{{\ttfamily 2304.10564}}].

\bibitem{Kosterlitz76}
J.M.~Kosterlitz, \emph{Phase transitions in long-range ferromagnetic chains},
  \href{https://doi.org/10.1103/PhysRevLett.37.1577}{\emph{Phys. Rev. Lett.}
  {\bfseries 37} (1976) 1577}.

\bibitem{Ising25}
E.~Ising, \emph{Beitrag zur theorie des ferromagnetismus},
  \href{https://doi.org/10.1007/BF02980577}{\emph{Zeitschrift f{\"u}r Physik}
  {\bfseries 31} (1925) 253}.

\bibitem{VojtaIsing1}
T.~Vojta, \emph{Disorder-induced rounding of certain quantum phase
  transitions},
  \href{https://doi.org/10.1103/PhysRevLett.90.107202}{\emph{Phys. Rev. Lett.}
  {\bfseries 90} (2003) 107202}.

\bibitem{VojtaIsing2}
J.A.~Hoyos and T.~Vojta, \emph{Theory of smeared quantum phase transitions},
  \href{https://doi.org/10.1103/PhysRevLett.100.240601}{\emph{Phys. Rev. Lett.}
  {\bfseries 100} (2008) 240601}.

\bibitem{Adrian08}
A.~{Del Maestro}, B.~{Rosenow}, M.~{M{\"u}ller} and S.~{Sachdev},
  \emph{{Infinite Randomness Fixed Point of the Superconductor-Metal Quantum
  Phase Transition}},
  \href{https://doi.org/10.1103/PhysRevLett.101.035701}{\emph{Phys. Rev. Lett.}
  {\bfseries 101} (2008) 035701}
  [\href{https://arxiv.org/abs/0802.3900}{{\ttfamily 0802.3900}}].

\bibitem{DasguptaMa}
C.~Dasgupta and S.-k.~Ma, \emph{{Low-temperature properties of the random
  Heisenberg antiferromagnetic chain}},
  \href{https://doi.org/10.1103/PhysRevB.22.1305}{\emph{Phys. Rev. B}
  {\bfseries 22} (1980) 1305}.

\bibitem{Wang:2019bpd}
Y.~Wang, \emph{{Solvable Strong-coupling Quantum Dot Model with a
  Non-Fermi-liquid Pairing Transition}},
  \href{https://doi.org/10.1103/PhysRevLett.124.017002}{\emph{Phys. Rev. Lett.}
  {\bfseries 124} (2020) 017002}
  [\href{https://arxiv.org/abs/1904.07240}{{\ttfamily 1904.07240}}].

\bibitem{Esterlis:2019ola}
I.~Esterlis and J.~Schmalian, \emph{{Cooper pairing of incoherent electrons: an
  electron-phonon version of the Sachdev-Ye-Kitaev model}},
  \href{https://doi.org/10.1103/PhysRevB.100.115132}{\emph{Phys. Rev. B}
  {\bfseries 100} (2019) 115132}
  [\href{https://arxiv.org/abs/1906.04747}{{\ttfamily 1906.04747}}].

\bibitem{Foster_22}
T.C.~{Wu}, Y.~{Liao} and M.S.~{Foster}, \emph{{Quantum interference of
  hydrodynamic modes in a dirty marginal Fermi liquid}},
  \href{https://doi.org/10.1103/PhysRevB.106.155108}{\emph{Phys. Rev. B}
  {\bfseries 106} (2022) 155108}
  [\href{https://arxiv.org/abs/2206.01762}{{\ttfamily 2206.01762}}].

\bibitem{Ilya1}
I.~{Esterlis}, H.~{Guo}, A.A.~{Patel} and S.~{Sachdev}, \emph{{Large-N theory
  of critical Fermi surfaces}},
  \href{https://doi.org/10.1103/PhysRevB.103.235129}{\emph{Phys. Rev. B}
  {\bfseries 103} (2021) 235129}
  [\href{https://arxiv.org/abs/2103.08615}{{\ttfamily 2103.08615}}].

\bibitem{Patel:2018mag}
A.A.~Patel, J.~McGreevy, D.P.~Arovas and S.~Sachdev, \emph{Magnetotransport in
  a model of a disordered strange metal},
  \href{https://doi.org/10.1103/PhysRevX.8.021049}{\emph{Phys. Rev. X}
  {\bfseries 8} (2018) 021049}.

\bibitem{LuntsHMC}
P.~Lunts, M.S.~Albergo and M.~Lindsey, \emph{{Non-Hertz-Millis scaling of the
  antiferromagnetic quantum critical metal via scalable Hybrid Monte Carlo}},
  \href{https://doi.org/10.1038/s41467-023-37686-4}{\emph{Nature
  Communications} {\bfseries 14} (2023) 2547}.

\bibitem{Hussey_FeSe}
S.~Licciardello, J.~Buhot, J.~Lu, J.~Ayres, S.~Kasahara, Y.~Matsuda et~al.,
  \emph{Electrical resistivity across a nematic quantum critical point},
  {\emph{Nature} {\bfseries 567} (2019) 213}.

\bibitem{Wolynes-rev}
M.D.~Ediger, M.~Gruebele, V.~Lubchenko and P.G.~Wolynes, \emph{Glass dynamics
  deep in the energy landscape},
  \href{https://doi.org/10.1021/acs.jpcb.1c01739}{\emph{The Journal of Physical
  Chemistry B} {\bfseries 125} (2021) 9052}.

\bibitem{Gruebele19}
H.A.~Nguyen, C.~Liao, A.~Wallum, J.~Lyding and M.~Gruebele, \emph{{Multi-scale
  dynamics at the glassy silica surface}},
  \href{https://doi.org/10.1063/1.5123228}{\emph{The Journal of Chemical
  Physics} {\bfseries 151} (2019) 174502}.

\bibitem{Yacoby20}
M.J.H.~{Ku}, T.X.~{Zhou}, Q.~{Li}, Y.J.~{Shin}, J.K.~{Shi}, C.~{Burch} et~al.,
  \emph{{Imaging viscous flow of the Dirac fluid in graphene}},
  \href{https://doi.org/10.1038/s41586-020-2507-2}{\emph{Nature} {\bfseries
  583} (2020) 537} [\href{https://arxiv.org/abs/1905.10791}{{\ttfamily
  1905.10791}}].

\end{thebibliography}\endgroup
\newpage


\section*{Supplementary material}

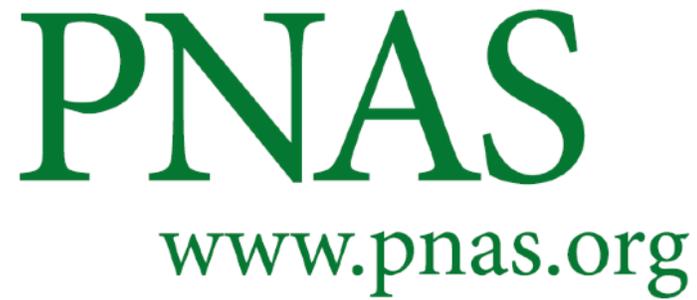

**Supplementary Information for**

**Localization of overdamped bosonic modes and transport in strange metals**

**Aavishkar A. Patel, Peter Lunts, and Subir Sachdev**

**Corresponding author:Aavishkar A. Patel**
**E-mail: apatel@flatironinstitute.org**

**This PDF file includes:**

Supplementary text
Figs. S1 to S7

# Supporting Information Text

## 1. Off-critical results

This section presents numerical results on the overdamped boson model in Eqs. [3] and [4] away from the critical point $\lambda = \lambda_c = -0.447$.

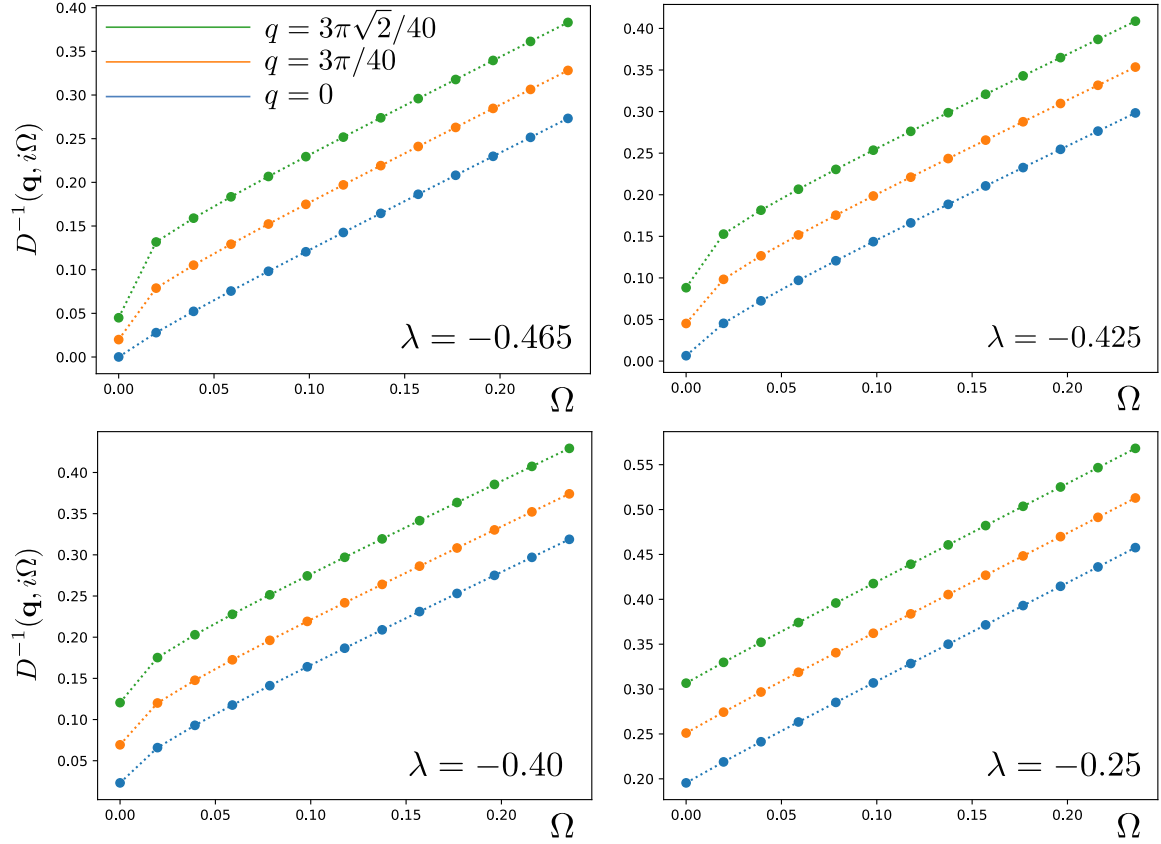

**Fig. S1.** The inverse boson propagator as a function of imaginary frequency, supplementing the critical point result in Fig. 2a. These results were obtained for a single disorder sample at $\beta = 320$.

The frequency dependence of the inverse boson propagator is in Fig. S1, to be compared with the critical point results in Fig. 2a. The frequency dependence is linear, apart from the low frequency regime for $\lambda \lesssim \lambda_c$, where deviations occur from the contribution of localized modes.

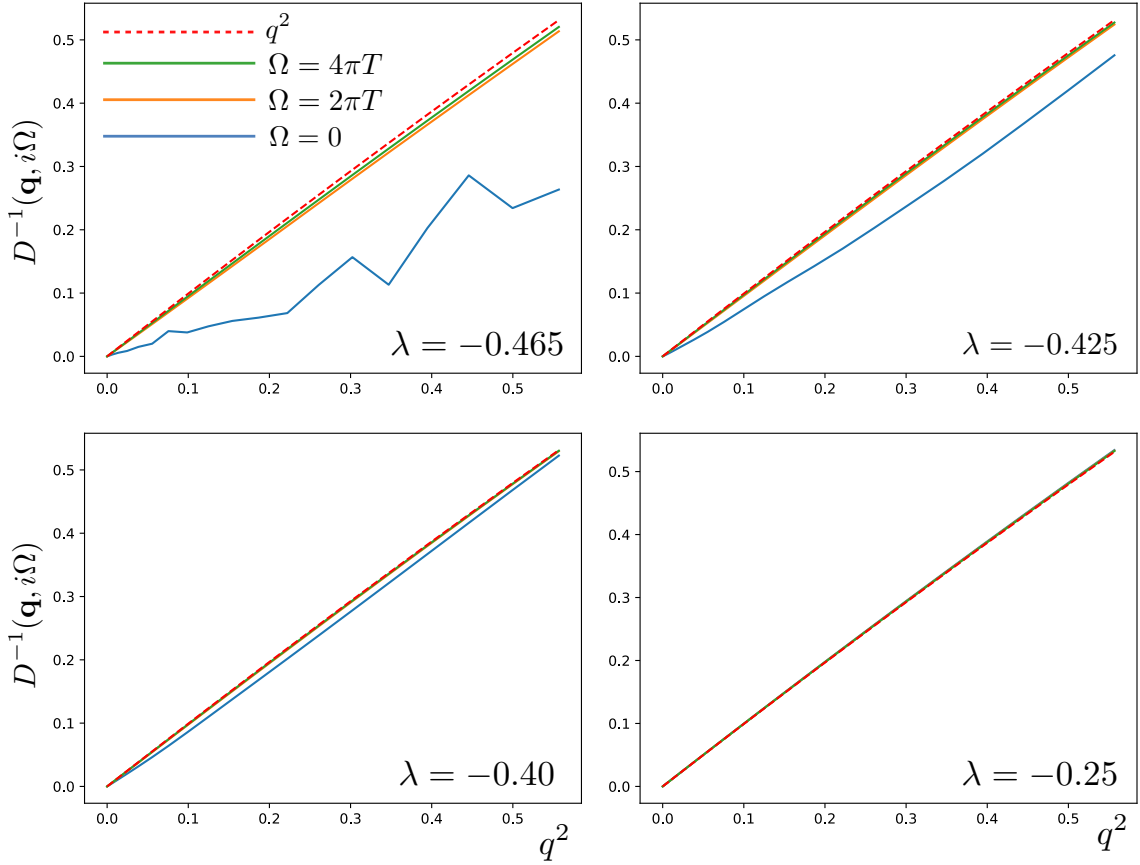

**Fig. S2.** The inverse boson propagator as a function of momentum squared, supplementing the critical point result in Fig. 2b. These results were obtained for a single disorder sample at $\beta = 320$.

The momentum dependence of the inverse boson propagator is in Fig. S2, to be compared with the critical point results in Fig. 2b. The dependence on $q^2$ is linear, apart from the $\Omega = 0$ Matsubara frequency for $\lambda \lesssim \lambda_c$, where deviations occur from the contribution of localized modes.

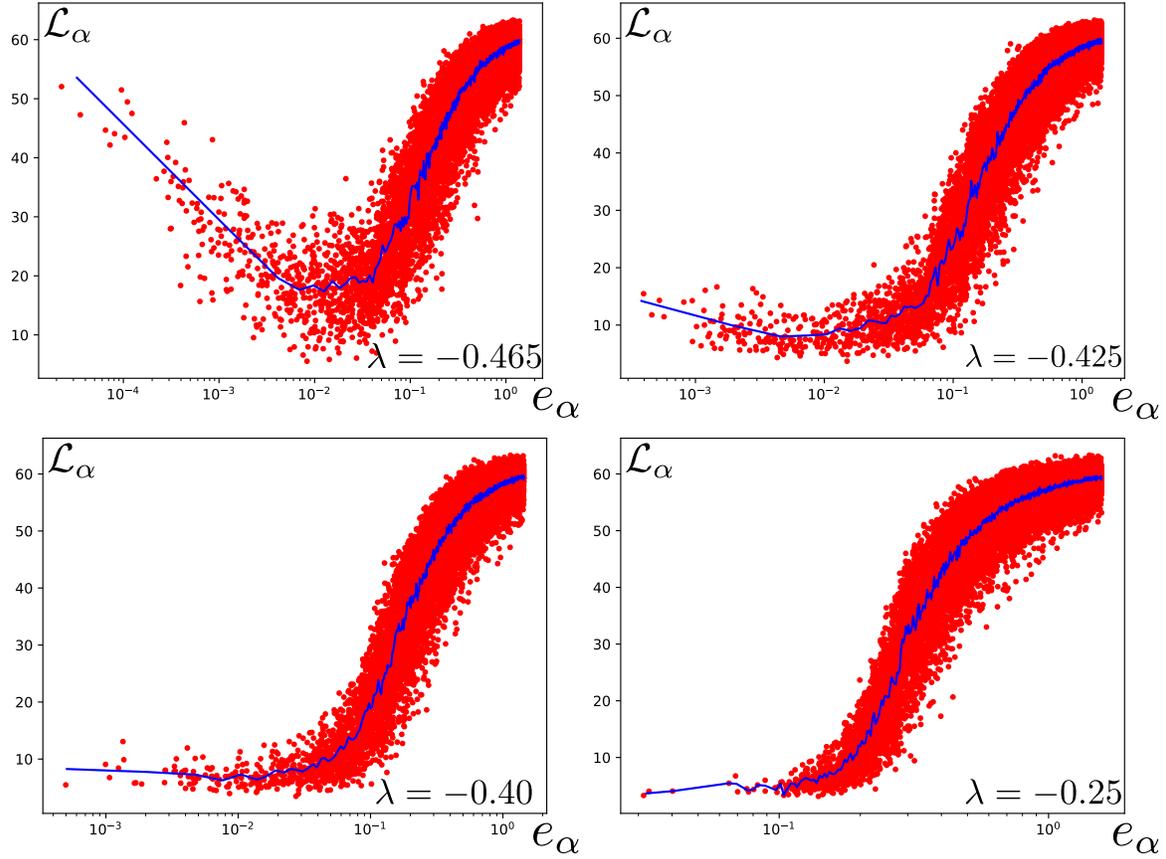

**Fig. S3.** Localization lengths away from the critical point, supplementing the critical point results in Fig. 3a. These results were obtained at $\beta = 320$ and combine 10 disorder samples.

Fig. S3 shows the localization of the bosonic eigenmodes, as in Fig. 3a. The non-monotonic behavior with decreasing energy is visible only for $\lambda \lesssim \lambda_c$.

## 2. Fermion self energy

We provide the analytical expression for the Matsubara frequency summation in the computation of the fermion self energy $\Sigma(i\omega)$ in Eq. [12]:

$$\Sigma(i\omega) = -i\pi g'^2 \mathcal{N}_0 \frac{T}{L^2} \sum_{\alpha,\Omega} \frac{\mathrm{sgn}(\omega+\Omega)}{\gamma|\Omega| + \Omega^2/c^2 + e_\alpha}$$

$$= -\sum_\alpha \frac{ig'^2 \mathrm{sgn}(\omega)\mathcal{N}_0}{e_\alpha (4e_\alpha - c^2\gamma^2)} \left[ \pi T \left( c^2\gamma^2 - 4e_\alpha \right) + ce_\alpha \sqrt{c^2\gamma^2 - 4e_\alpha} \left\{ -\mathcal{G}\left( \frac{|\omega|}{2\pi T} + \frac{c\left(c\gamma - \sqrt{c^2\gamma^2 - 4e_\alpha}\right)}{4\pi T} + \frac{1}{2} \right) \right.\right.$$

$$\left.\left. + \mathcal{G}\left( \frac{|\omega|}{2\pi T} + \frac{c\left(c\gamma + \sqrt{c^2\gamma^2 - 4e_\alpha}\right)}{4\pi T} + \frac{1}{2} \right) + \mathcal{G}\left( \frac{c\left(c\gamma - \sqrt{c^2\gamma^2 - 4e_\alpha}\right)}{4\pi T} \right) - \mathcal{G}\left( \frac{c\left(c\gamma + \sqrt{c^2\gamma^2 - 4e_\alpha}\right)}{4\pi T} \right) \right\} \right],$$

$$\Sigma^R(\omega) = -\sum_\alpha \frac{ig'^2 \mathcal{N}_0}{e_\alpha (4e_\alpha - c^2\gamma^2)} \left[ \pi T \left( c^2\gamma^2 - 4e_\alpha \right) + ce_\alpha \sqrt{c^2\gamma^2 - 4e_\alpha} \left\{ -\mathcal{G}\left( \frac{-i\omega}{2\pi T} + \frac{c\left(c\gamma - \sqrt{c^2\gamma^2 - 4e_\alpha}\right)}{4\pi T} + \frac{1}{2} \right) \right.\right.$$

$$\left.\left. + \mathcal{G}\left( \frac{-i\omega}{2\pi T} + \frac{c\left(c\gamma + \sqrt{c^2\gamma^2 - 4e_\alpha}\right)}{4\pi T} + \frac{1}{2} \right) + \mathcal{G}\left( \frac{c\left(c\gamma - \sqrt{c^2\gamma^2 - 4e_\alpha}\right)}{4\pi T} \right) - \mathcal{G}\left( \frac{c\left(c\gamma + \sqrt{c^2\gamma^2 - 4e_\alpha}\right)}{4\pi T} \right) \right\} \right],$$

[S1]

were $\mathcal{G}$ is the digamma function.

## 3. $M = 1$ Ising scalar theory

In this section we show the properties of the $M = 1$ bosons. The diagonal part of the inverse propagator as a function of frequency and momentum squared across the critical region is shown in Figs. S4 and S5.

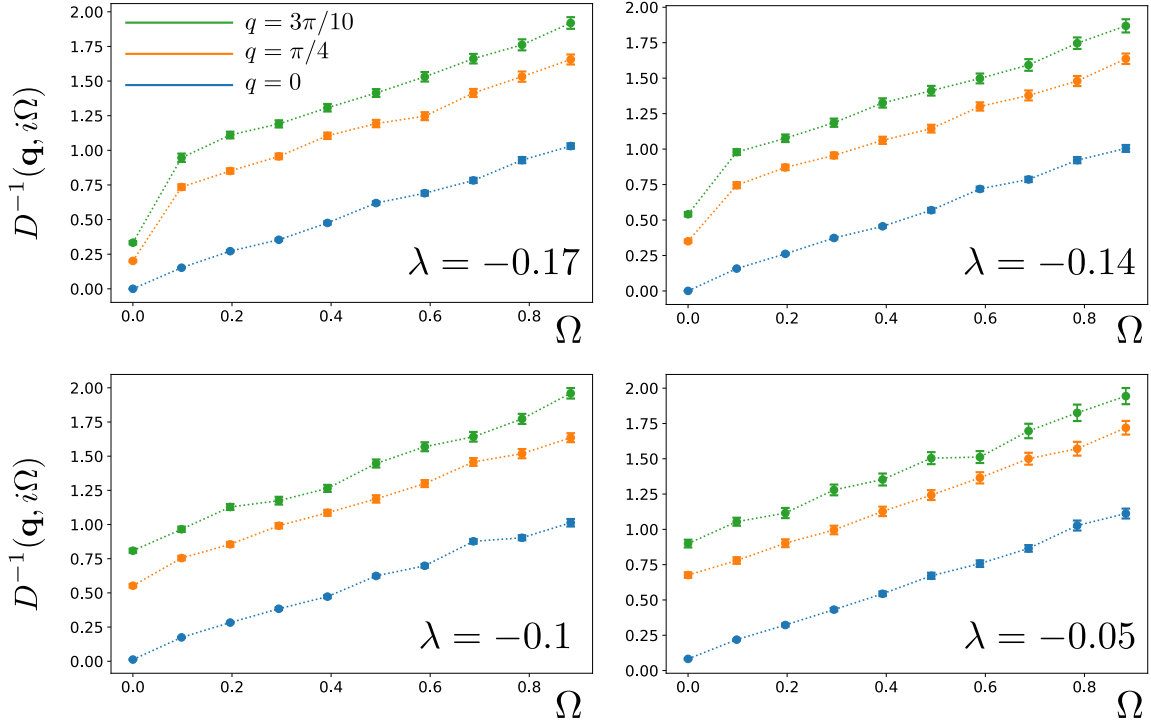

**Fig. S4.** The inverse propagator as a function of imaginary frequency, going across the smeared critical region. These results were obtained at $\beta = 64$, $L = 80$ and combine between $40$ ($\lambda = -0.05$) and $200$ disorder samples.

We can see that the as a function of frequency, the inverse propagator is basically linear for all non-zero frequencies, just as in the $M \geq 2$ case shown in Fig. S1.

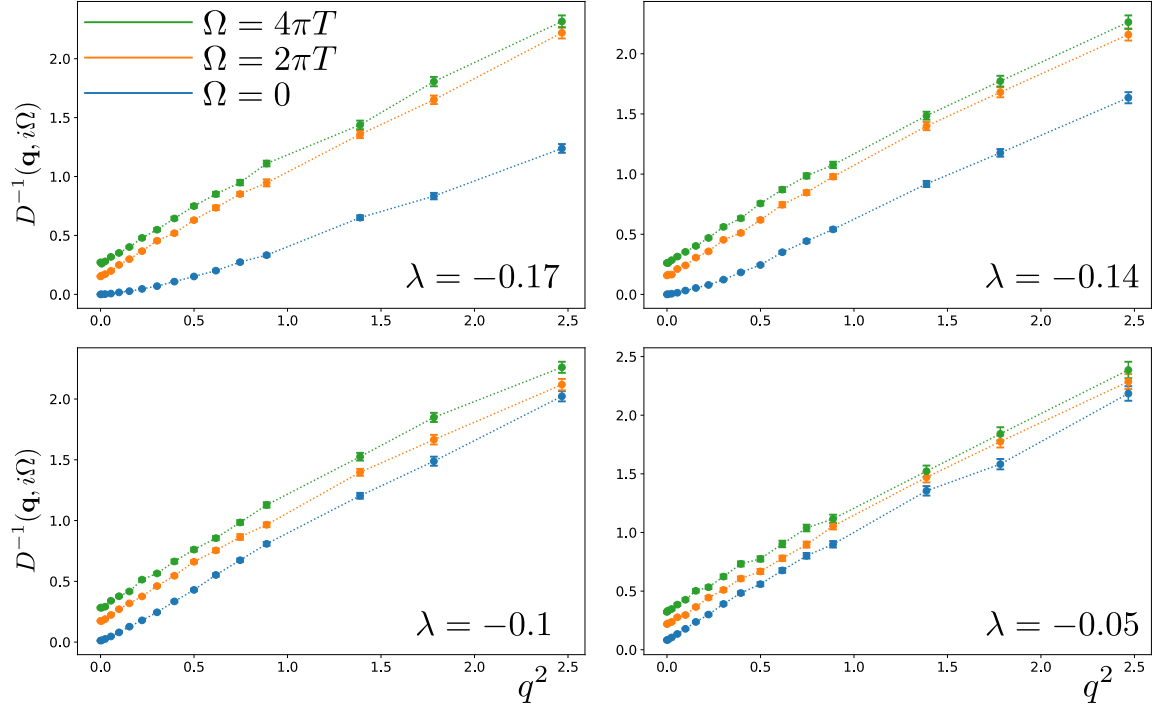

**Fig. S5.** The inverse propagator as a function of momentum squared, going across the smeared critical region. These results were obtained at $\beta = 64$, $L = 80$ and combine between 40 ($\lambda = -0.05$) and 200 disorder samples.

As a function of momentum squared, the inverse propagator again looks very similar to the $M \geq 2$ case shown in Fig. S2. The main difference is that there is no sharp transition beyond which only the zero-frequency propagator acquires a strong disorder-sample-dependence: here all the curves are strongly fluctuating as a function of disorder samples. The disorder fluctuations decrease gradually as we move to larger values of $\lambda$, but one disorder configuration is not enough to give a smooth curve for any parameter value we looked at.

We also plot the off-diagonal momentum components of $D(\boldsymbol{q}_1 = 0, \boldsymbol{q}_2, i\Omega)$ in Fig. S6 for a single disorder sample. Unlike the $M \geq 2$ case, all Matsubara frequencies show a significant off-diagonal component, indicating a much larger disorder variance.

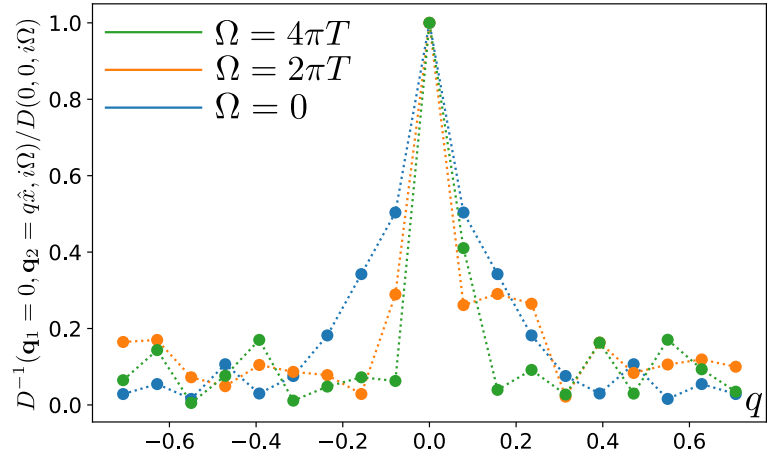

**Fig. S6.** Momentum off-diagonal components of $D(\boldsymbol{q}_1 = 0, \boldsymbol{q}_2, i\Omega)$ for different Matsubara frequencies, in the middle of the smeared critical region ($\lambda = -0.14$). This is for a single disorder sample at $\beta = 64$, $L = 80$. All Matsubara frequencies show a significant off-diagonal component.

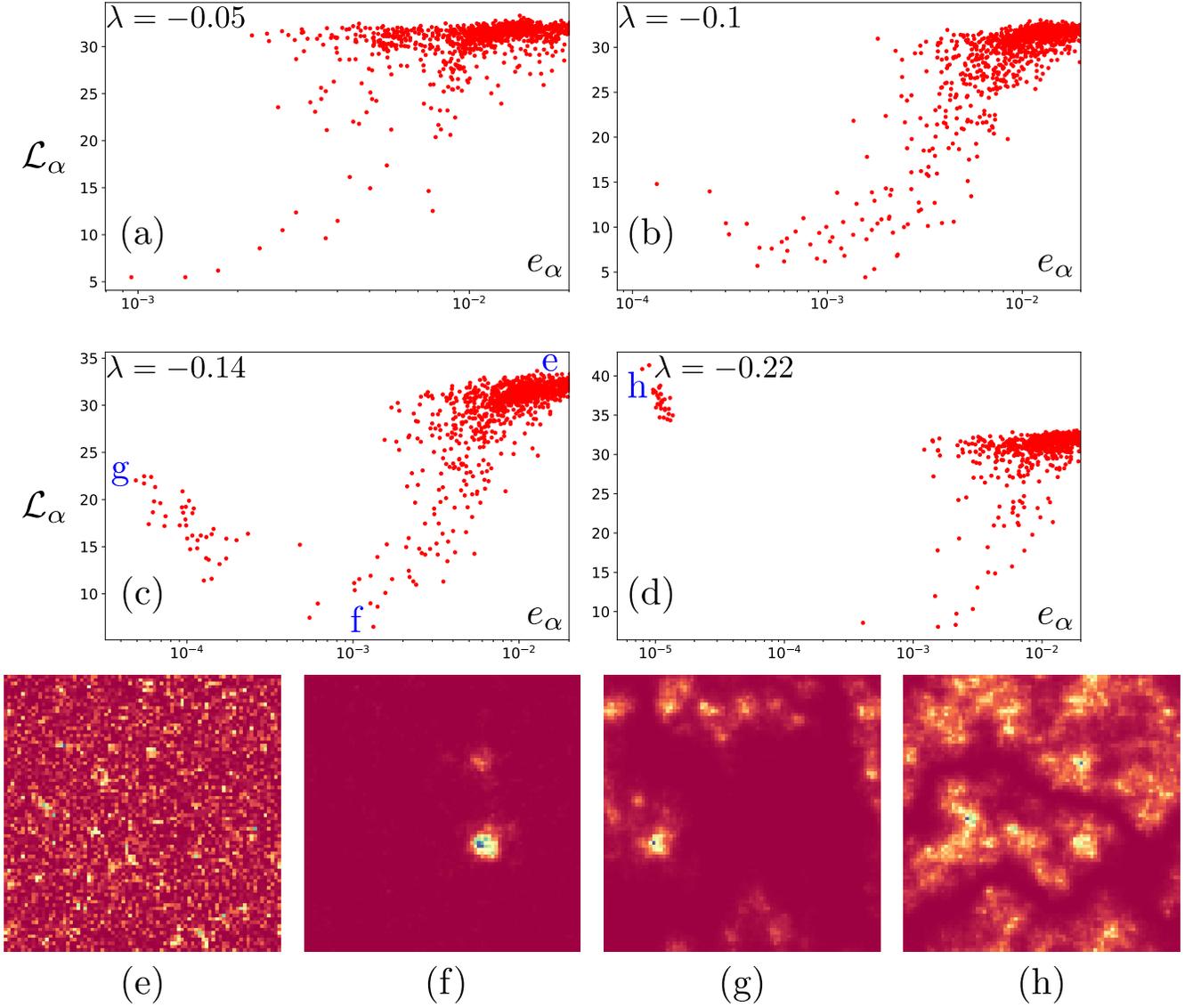

**Fig. S7.** (a) - (d): Localization lengths across the smeared critical region. These results were obtained at $\beta = 64$, $L = 80$ and combine 40 disorder samples. These results should be compared with Fig. S3 (the number of disorder samples is chosen to match the density for the two systems of different sizes). (e) - (g): three different wavefunction probabilities at $\lambda = -0.14$, plotted in real space. As expected, the three types of states correspond to (e) plane waves, (f) localized modes, (g)-(h) delocalized non-translation-invariant modes for $\lambda = -0.14$ and $\lambda = -0.22$.

We show the localization lengths of the eigenmodes of the zero-frequency propagator in Fig. S7. We can see that the delocalized non-plane-wave modes start to appear at the far edge of the smeared critical region, relatively far away from the peak of the susceptibility. Importantly, unlike the $M \geq 2$ case, as $\lambda$ is lowered, a gap opens up in the density of states that separates these states and the others (localized and plane-wave states). We attribute these states to the non-fluctuating ordered puddles. Interestingly, deeper in the ordered phase, these puddles can become more delocalized that the plane-wave states, and can essentially take up the entire size of the system (or half of it, given the periodic boundary conditions.), as shown in Fig. S7h.



\end{document}